\documentclass[11pt,preprint]{aastex}
\newcommand{\beq}{\begin{equation}}
\newcommand{\eeq}{\end{equation}}
\newcommand{\beqa}{\begin{eqnarray}}
\newcommand{\eeqa}{\end{eqnarray}}
\slugcomment{First draft, \date}
\shorttitle{DES 2010/001, docDB \#nnnn}
\shortauthors{Rossetto et al.}
\begin{document}
\title{The Dark Energy Survey: Prospects for Resolved Stellar Populations}

\author
{ Bruno M. Rossetto\altaffilmark{1,2}, Bas\'\i lio X. Santiago \altaffilmark{3,2}, L\'eo Girardi \altaffilmark{4,2}, Julio I.B. Camargo \altaffilmark{1,2}, Eduardo Balbinot \altaffilmark{3,2}, Luiz N. da Costa \altaffilmark{1,2}, Brian Yanny \altaffilmark{5}, Marcio A.G. Maia \altaffilmark{1,2}, Martin Makler\altaffilmark{6,2}, Ricardo L.C. Ogando \altaffilmark{1,2}, Paulo S. Pellegrini \altaffilmark{1,2}, Beatriz Ramos \altaffilmark{1,2}, Fernando de Simoni \altaffilmark{1,2}, R. Armstrong,\altaffilmark{10}, E. Bertin,\altaffilmark{8}, S. Desai,\altaffilmark{9,10}, N. Kuropatkin,\altaffilmark{7}, H. Lin,\altaffilmark{7}, J. J. Mohr,\altaffilmark{12,13,11}, D. L. Tucker,\altaffilmark{7} }
\altaffiltext{1} {Observat\'orio Nacional, Rua Gal. Jos\'e Cristino 77, Rio de Janeiro, RJ - 22460-040, Brazil}
\email{rossetto@linea.gov.br}
\altaffiltext{2} {Laborat\'orio Interinstitucional de e-Astronomia - LIneA, Rua Gal. Jos\'e Cristino 77, Rio de Janeiro, RJ - 20921-400, Brazil}
\altaffiltext{3} {Instituto de F\'\i sica, UFRGS, Caixa Postal 15051, Porto Alegre, RS - 91501-970, Brazil}
\altaffiltext{4} {Osservatorio Astronomico di Padova - INAF, Vicolo dell'Osservatorio 5, I - 35122 Padova, Italy}
\altaffiltext{5} {Fermi National Laboratory, P.O. Box 500, Batavia, IL 60510-5011, USA}
\altaffiltext{6} {Centro Brasileiro de Pesquisas F\'\i sicas, Rua Dr. Xavier Sigaud 150, Rio de Janeiro, RJ - 22290-180, Brazil}
\altaffiltext{7}{Fermi National Accelerator Laboratory, P.O. Box 500, Batavia, IL 60510}
\altaffiltext{8}{Institut d'Astrophysique de Paris, UMR 7095 CNRS, Universit\'e Pierre et Marie Curie, 98 bis boulevard Arago, F-75014 Paris, France}
\altaffiltext{9}{Department of Astronomy, University of Illinois, 1002 West Green Street, Urbana, IL 61801}
\altaffiltext{10}{National Center for Supercomputing Applications, University of Illinois, 1205 West Clark Street, Urbanan, IL 61801}
\altaffiltext{11}{Max-Planck-Institut f\"{u}r extraterrestrische Physik, Giessenbachstr.\ 85748 Garching, Germany}
\altaffiltext{12}{Department of Physics, Ludwig-Maximilians-Universit\"{a}t, Scheinerstr.\ 1, 81679 M\"{u}nchen, Germany}
\altaffiltext{13}{Excellence Cluster Universe, Boltzmannstr.\ 2, 85748 Garching, Germany} 
%
%
%
%
\newpage
\begin{abstract}
Wide angle and deep surveys, regardless of their primary purpose, always sample a large number of stars in the Galaxy and in its satellite system. We here make a forecast of the expected stellar sample resulting from the Dark Energy Survey and the perspectives that it will open for studies
of Galactic structure and resolved stellar populations in general. An estimated $1.2\times10^8$ stars will be sampled in DES $grizY$ filters in the southern equatorial hemisphere. This roughly corresponds to 20\% of all DES sources. Most of these stars belong to the stellar thick disk and halo of the Galaxy. DES will probe low-mass stellar and sub-stellar objects at depths from 3 to 8 times larger than SDSS. The faint end of the main-sequence will be densely sampled beyond 10 kpc. The slope of the low mass end of the stellar IMF will be constrained to within a few hundredth dex, even in the thick disk and halo. In the sub-stellar mass regime, the IMF slope will be potentially constrained to within $d \log \phi(m) / d \log m \simeq 0.1$. About $3\times10^4$ brown dwarf and at least $7.6\times10^5$ white dwarf candidates will be selected, the latter embeded into the thick disk and halo, for future follow-up. The stellar halo flattening will also be constrained to within a few percent. DES will probe the main sequence of new Milky Way satellites and halo clusters for distances out to $\simeq 120$kpc, therefore yielding stellar surface density contrasts $1.6-1.7$ times larger than those attainable with SDSS. It will also allow detection of these objects in the far reaches of the stellar halo, substantially increasing the number and quality of probes to the Galactic potential. Combined with northern samples, such as the SDSS, the DES stellar sample will yield constraints on the structure and stellar populations of Galactic components in unprecedented detail. In particular, the combined sample from both hemispheres will allow detailed studies of halo and thick disk asymmetries and triaxiality.
\end{abstract}


%
\section{Introduction}
\label{intro}
Wide angle photometric surveys have been of enormous value in studies of the structure and history of formation of our galaxy. Surveys such as the Two Micron All-Sky Survey \citep[2MASS;][]{Skrutskie1997} and the Sloan Digital Sky Survey \citep[SDSS;][]{York2000} have brought a substantial increase in the size and depth of homogeneous stellar samples. As a consequence, the structural parameters of the
Galaxy have been determined with unmatched precision \citep[e.g.;][]{Juric2008, Reyle09}. The census of low mass and low luminosity stars and sub-stellar objects has gone through a major boost \citep[e.g.;][]{Zhang09, Kilic2006}. These surveys have also revealed the complexity of substructure in the Galactic halo \citep[e.g.;][]{Kop2008,Walsh2009,RochaPinto04,Sharma10}.

A new generation of wide angle and deep surveys is underway. In the infra-red, the UKIRT Infra-Red Deep Sky Survey \cite[UKIDSS,][]{Lawrence2007} is covering 7500 deg$^2$ of the northern celestial hemisphere in $JHK$, reaching about 3 magnitudes deeper than 2MASS. Its ability to see through the high dust extinction regions is allowing new discoveries in previously unexplored Galactic depths \citep[e.g.;][]{Lucas10}. The Visible and Infrared Survey Telescope for Astronomy (VISTA) will carry out several surveys in the near infra-red $ZYJHKs$ bands during the upcoming years. In particular, the Vista Hemisphere Survey (VHS) will cover the entire southern sky in $J$ and $K_s$, and a smaller region in $H$, with limiting magnitudes reaching $J=21.2$, $H=20.6$, and $K_s=20.0$. The VISTA Magellanic Clouds Survey (VMC) will sample the areas in and around these two satellites at much fainter levels than 2MASS \citep{Cioni2008}.

The Dark Energy Survey (DES) is a survey planned to start in 2011 \citep[][]{Depoy2008,Mohr2008}. DES will cover a continuous area of 5,000 deg$^2$ of the southern galactic cap, in five ($grizY$) filters, to limiting magnitudes much fainter than those of SDSS over a comparable area, thus probing a much larger volume. It will use a new large field of view camera, with a mosaic of 64 CCDs of considerable efficiency in the near-infrared. The combination of a large area, depth and red sensitivity of the survey makes it of particular interest in the search for cool stars in the Galaxy.

Also in the optical region, the SkyMapper Southern Sky Survey \citep[][]{Murphy2009} will cover the entire Southern Hemisphere on 6 filters ($uvgriz$) but will be shallower than DES. On the other hand, addition of $u$-band data is particularly important to select several types of hot stars, from horizontal branch to white dwarfs \citep[][]{Sirko2004, Harris2003}.

Looking a little further into the future, when fully operational the Panoramic Survey Telescope and Rapid Response System (PanSTARSS) will consist of 4 mirrors of 1.8m in diameter each, simultaneously observing a $\simeq7$ deg$^2$ field of view. In a single lunar cycle, using dark time, it will be able to scan the entire northern sky about 3 times. Its main goal is to detect and monitor Solar System objects, especially those that can be potentially hazardous. On the other hand, its extremely large and multi-epoch database of stars and galaxies will have numerous other applications.

The Large Synoptic Survey Telescope (LSST) is an 8.5m telescope of extremely high \'etendue. With a $9.6$ deg$^2$ field of view, it will be capable of imaging half of the sky in 6 optical bands ($ugrizy$) once every few nights. One of its basic and unique strengths will be to discover and study
transient optical phenomena, produced from a variety of astrophysical sources. With its high observational rate, LSST will also be particularly suited to detect the non-sidereal motion of extremely faint Solar System objects. Finally, its massive and deep sampling of stellar and extra-galactic objects will allow research on several fields of astrophysics, from stellar to
cosmology \citep[][]{Ivezic08}.

The goal of the present paper is to investigate what are the possible contributions that DES will bring to the structure of our galaxy, to constrain key parameters describing the IMF and star formation history of different Galactic components, and to the identification of rare or faint stellar and
sub-stellar objects. Our starting point for this goal is a forecast of what the DES stellar sample will look like, based on the TRILEGAL galactic model presented by \citet{Girardi2005}. A preliminary forecast of DES stellar population studies was presented by \citet{Santiago2010}. In \S \ref{survey} we describe DES in more detail. In \S \ref{models}, we briefly review the Galaxy model used in our DES simulation and the choice of parameters adopted. We also present the expected spatial and photometric distributions within DES footprint for different stellar types and Galactic structural components.
In \S \ref{stargalsep} we deal with the critical issue of star/galaxy separation in a deep survey like DES. In \S \ref{stellar}, we discuss the DES characteristic sampling depths for low-mass and/or low-luminosity stars, including M stars, white dwarfs, and brown dwarfs. We also estimate the number of candidates of these objects to be sampled with DES under different assumptions for the stellar IMF at low-masses regime. In \S \ref{proper} we discuss the use of proper motions in association with DES photometric data as a tool to detect cool white dwarfs and to separate L and T dwarfs from high redshift QSOs. In \S \ref{halostruct}, we review recent results obtained for the Galactic structure, with emphasis on the thick disk and halo components. We also assess the constraints that DES will provide in determining the stellar shape of the halo and the prospects for searching for thick disk and halo asymmetries with the combined DES and SDSS samples. Finally, we discuss the perspectives brought by DES to the detection of new
Milky Way satellites, clusters and substructure in general. A summary of our results is presented in \S \ref{summary}.
\begin{figure*}
  \begin{minipage}{0.50\textwidth}
    \resizebox{\textwidth}{!}{\includegraphics{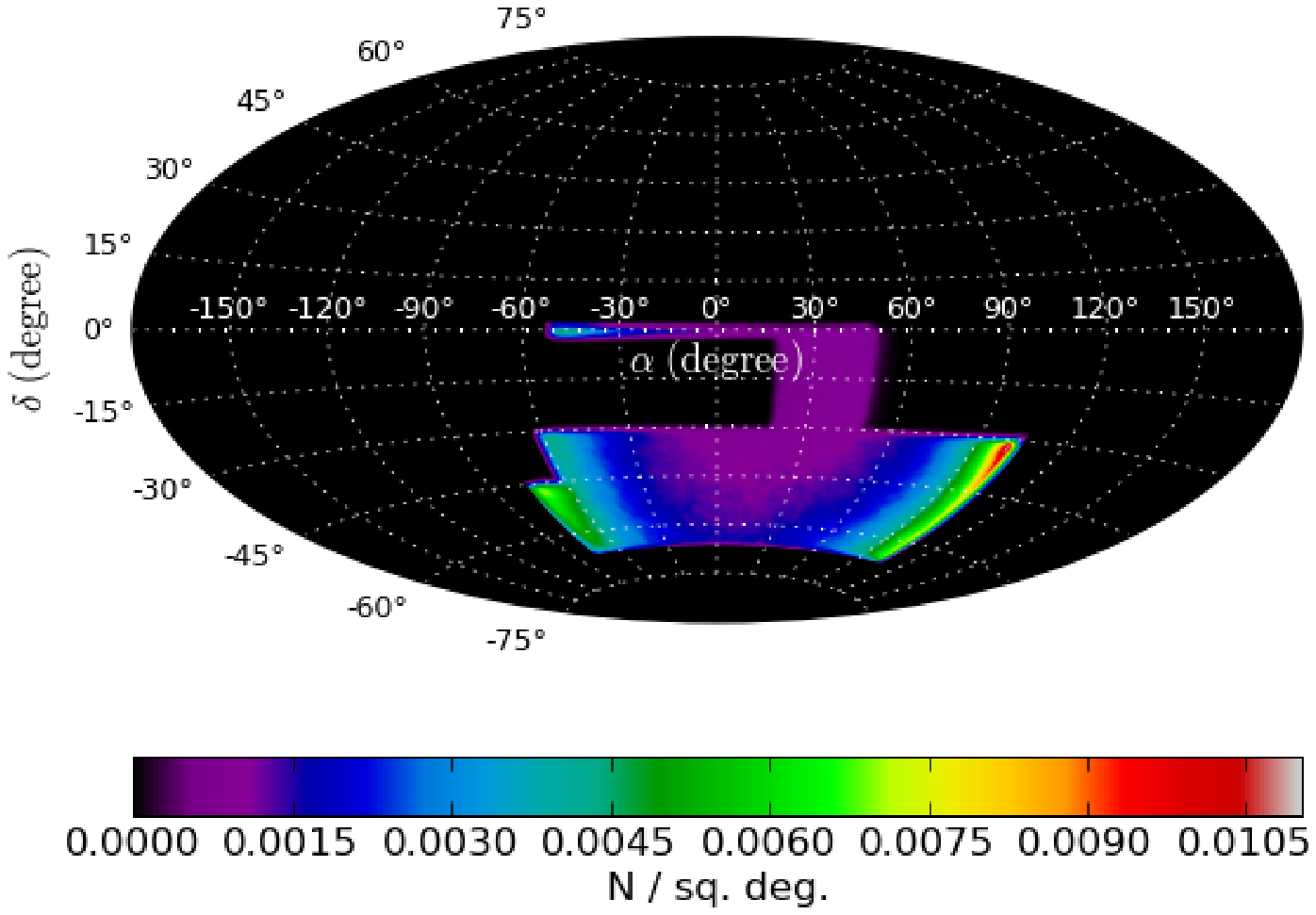}}
  \end{minipage}
  \begin{minipage}{0.50\textwidth}
    \resizebox{\textwidth}{!}{\includegraphics{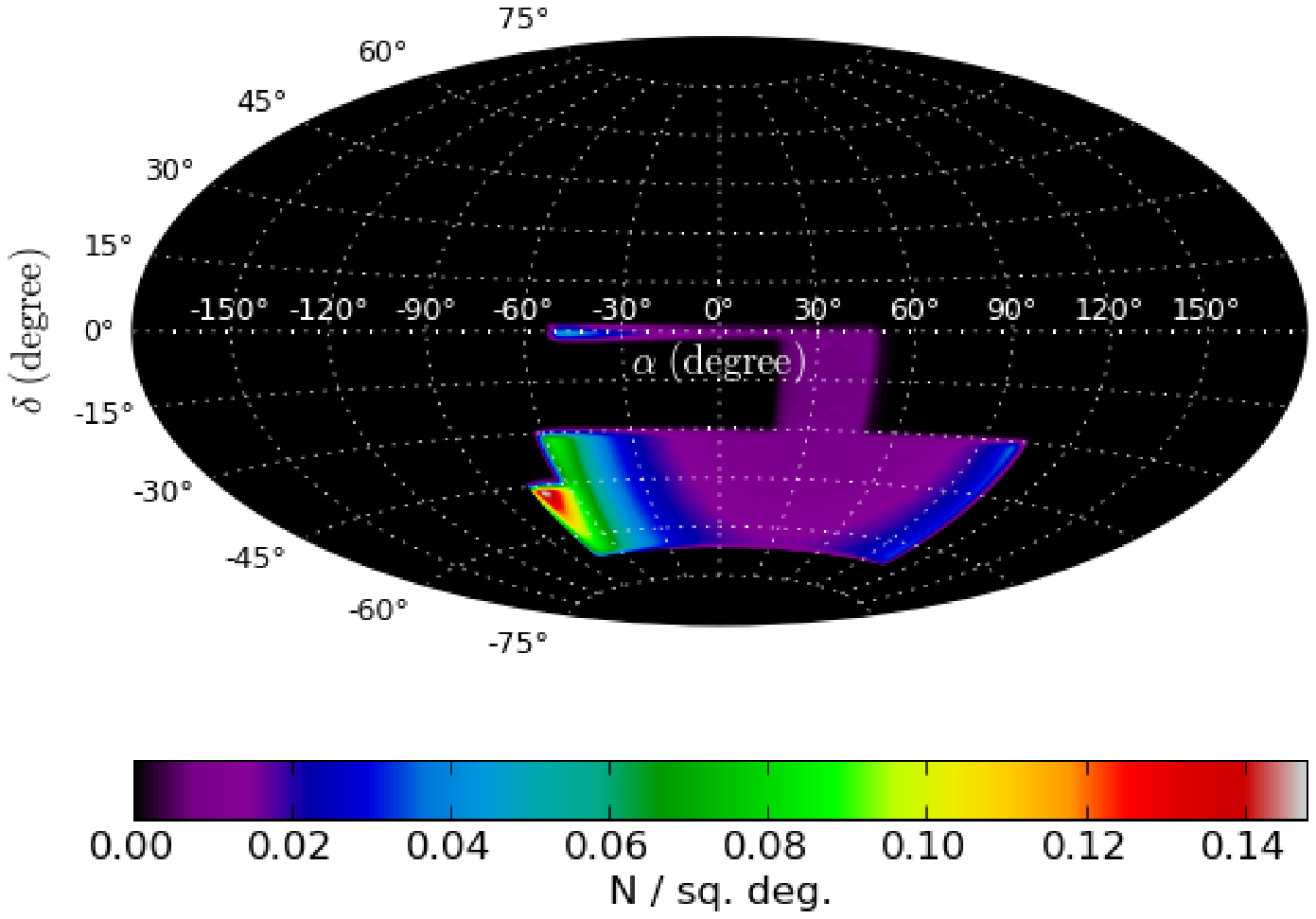}}
  \end{minipage}
\\
  \begin{minipage}{0.50\textwidth}
    \resizebox{\textwidth}{!}{\includegraphics{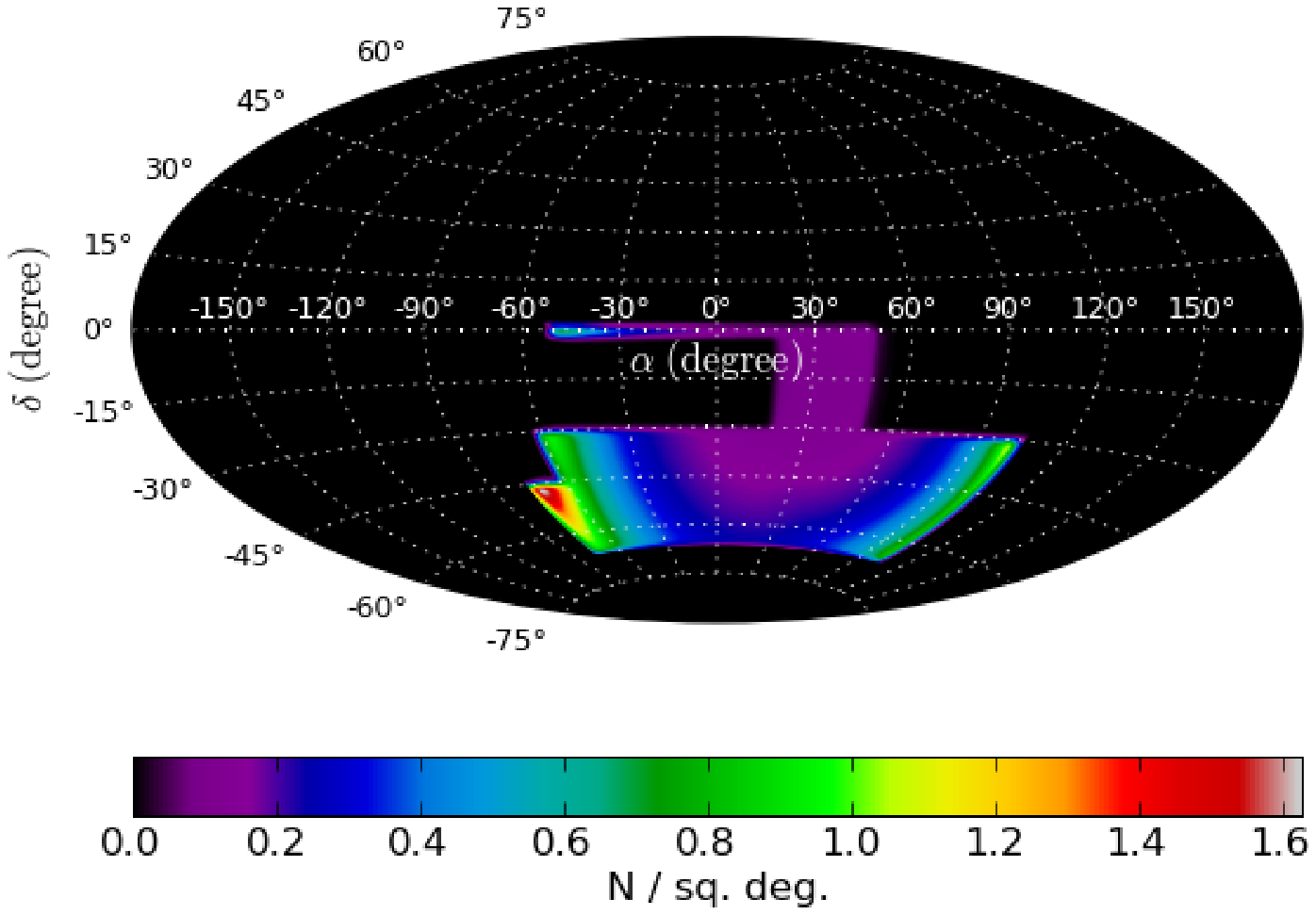}}
  \end{minipage}
  \begin{minipage}{0.50\textwidth}
    \resizebox{\textwidth}{!}{\includegraphics{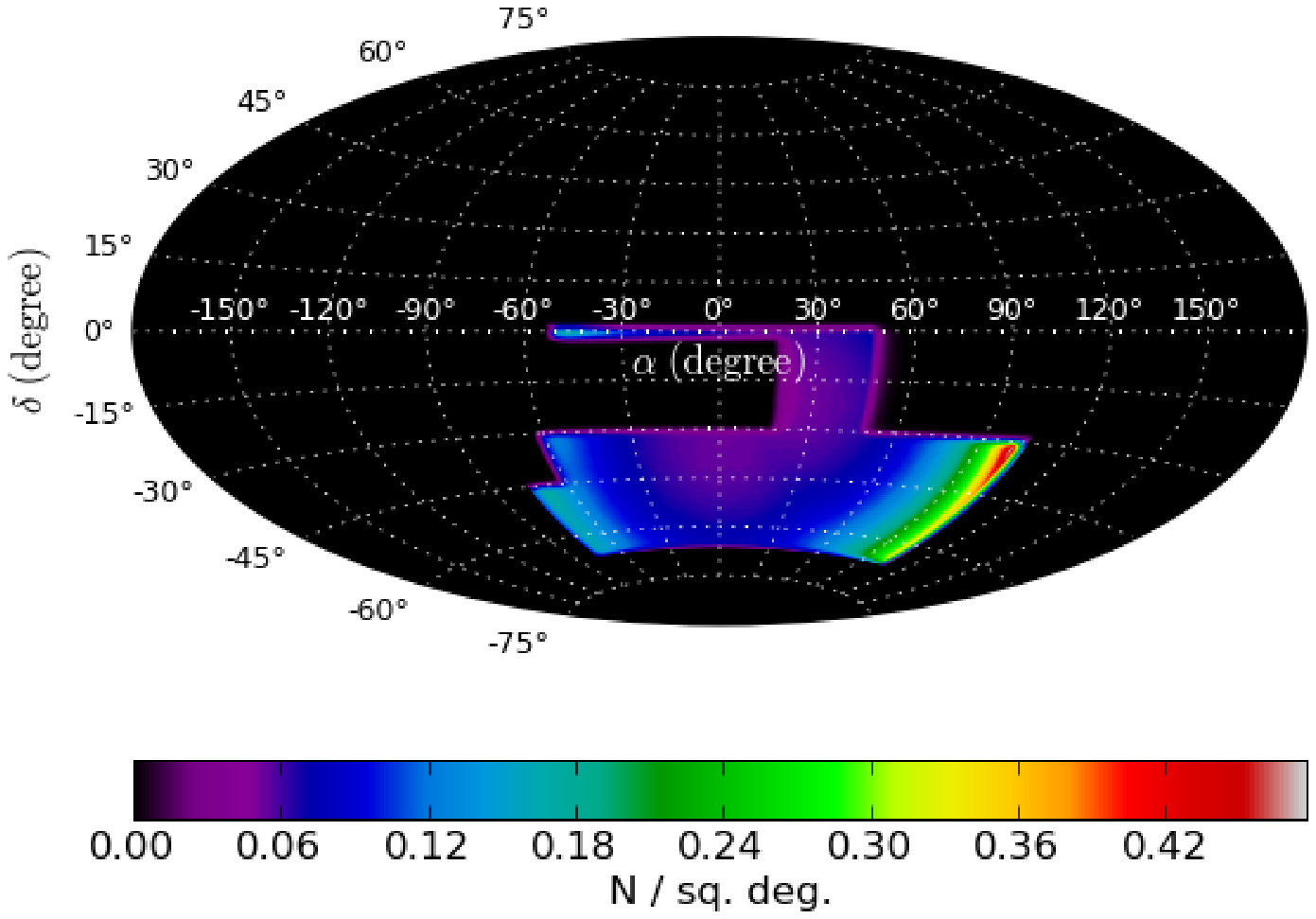}}
  \end{minipage}
  \caption{\small Spatial distribution of stars simulated within the DES area. The panels show stars of different spectral types: top left: O and B stars; top right: A and F; bottom left: G and K; bottom right: M and L.}
  \label{spat_dist}
\end{figure*}
\section{The Dark Energy Survey}
\label{survey}
During its planned 5 years of operation, DES will cover 5000 deg$^2$ on the sky in $grizY$ filters, mostly in directions with Galactic latitudes $\vert b \vert > 30^o$. A new large field camera, DECam, is currently under construction and will be installed in the primary focus of the 4m Blanco telescope, at Cerro Tololo Inter-American Observatory (CTIO) in order to carry out the survey. DECam is made up of 62 2kX4k red-sensitive CCDs recently developed at Lawrence Berkeley National Laboratory (LBNL). This large CCD mosaic will cover a field of view of 2.2 deg$^2$, yielding a pixel scale of 0.27 arcsec/pixel. Given the typical seeing at CTIO, ranging from 0.6 to 1.1 arcsec, this image scale will allow an efficient sampling of the point spread function and its variations across the field. The expected photometric calibration accuracy is of 2\% over the entire survey area.

DES primary purpose is to constrain the cosmological parameters, most specially the matter density parameter, $\Omega_m$, and the slope of the cosmic equation of state, $w$. As such, it was designed to include the sky region to be covered by the South Pole Telescope \citep[SPT,][]{Carlstrom2009}, in order to optically probe a potentially unbiased sample of massive galaxy clusters to high redshifts, pre-selected with the Sunyaev-Zeldovich effect \citep[][]{Staniszewski2009}. In order to be placed in a similar photometric system as SDSS, DES will also cover SDSS's Stripe 82 plus a connecting region between this latter and the SPT.

\placetable{tabstruct}
\begin{table}
\begin{center}
\caption{\small TRILEGAL parameters used in this work.
\label{tabstruct}}
\resizebox{0.47\textwidth}{!} {
\begin{tabular}{lll}
\tableline\tableline
Parameter & Value & Meaning \\
\tableline
\multicolumn{3}{l}{Exponential thin disk} \\
$\Sigma_{thin}$ & $59~M_{\odot}~pc^{-2}$ & local mass surface density \\
$h_R$ & 2800~pc & thin disk scale length \\
$h_z$ & 95~pc & young thin disk sech$^2$ scale height \\
$h_z^{\rm dust}$ & 100~pc & dust scale height \\
\tableline
\multicolumn{3}{l}{Exponential thick disk} \\
$\rho_{thick}$ & $0.004~M_{\odot}~pc^{-3}$ & local mass density \\
$h_R$ & 2800~pc & thick disk scale length \\
$h_z$ & 800~pc & thick disk exponential scale height \\
\tableline
\multicolumn{3}{l}{Oblate Halo} \\
$\rho_{halo}$ & $1.5\times10^{-4}~M_{\odot}~pc^{-3}$ & local mass density \\
$r_e$ & 2800~pc & de Vaucouleurs radius \\
$c/a$ & 0.65 & axial ratio \\
\tableline
\multicolumn{3}{l}{Triaxial Bulge} \\
$\rho_{bulge}$ & $406~M_{\odot}~pc^{-3}$ & central mass density \\
$r_e$ & 2500~pc & exponential scale radius \\
$b/a$ & 0.68 & first axial ratio \\
$c/a$ & 0.31 & second axial ratio \\
$\theta_b$ & $15^o$ & long axis orientation \\
\tableline
\end{tabular}
}
\end{center}
\end{table}

Despite its cosmological motivations, about 20\% of the detected DES sources will be stars, in a total of $\simeq 10^8$, the majority of them located in the thick disk and halo of the Milky Way. The survey's aimed detection limits are $g=24.6$, $r=24.1$, $i=24.4$, $z=23.8$, and $Y=21.3$ at a $S/N=10$. These deep limits will allow a wide range of studies on the structure of the Galaxy and
its satellite system. A first step towards assessing these contributions in detail requires a full model of the DES stellar sample. We do that in the next section.

As mentioned in \S \ref{intro}, DES will be complemented in several ways by other surveys in the southern skies and on similar timescales. Most specially, it will provide $Y$ band data to the VISTA collaboration in exchange for $JHK$ data within the DES area. At shorter wavelengths, SkyMapper will cover the $u$-band, although at a shallower depth and coarser image scale. Finally, as mentioned earlier, combination of microwave SPT data and optical DES data will be very useful for galaxy clusters detection and analysis.

\section{Galactic Model}
\label{models}

In this paper we adopt the TRIdimensional modeL of thE GALaxy (TRILEGAL) to describe the structure of the Galaxy. The model provides a synthetic stellar sample for any direction of the Galaxy, including contributions from its 4 basic structural components: the thin and thick disks, the bulge and the stellar halo. The code uses the \citet{Girardi2002} database of stellar isochrones, complemented with additional models for brown dwarfs \citep[][]{Allard2000, Chabrier2000} and white dwarfs \citep[][]{Benvenuto1999, Finley1997, Homeier1998}. More details about this Galactic model are provided in \citet{Girardi2005}.

The parameters describing the various Galactic components were originally determined by comparing model predictions to a large variety of pencil beam observations (e.g., DMS, EIS-deep), shallow wide-area surveys (e.g., 2MASS) and also the local sample from Hipparcos catalogue. The model calibration has been extended to the bulge of our Galaxy by \citet{Vanhollebeke2009}.

\begin{figure}
\resizebox{\hsize}{!}{\includegraphics{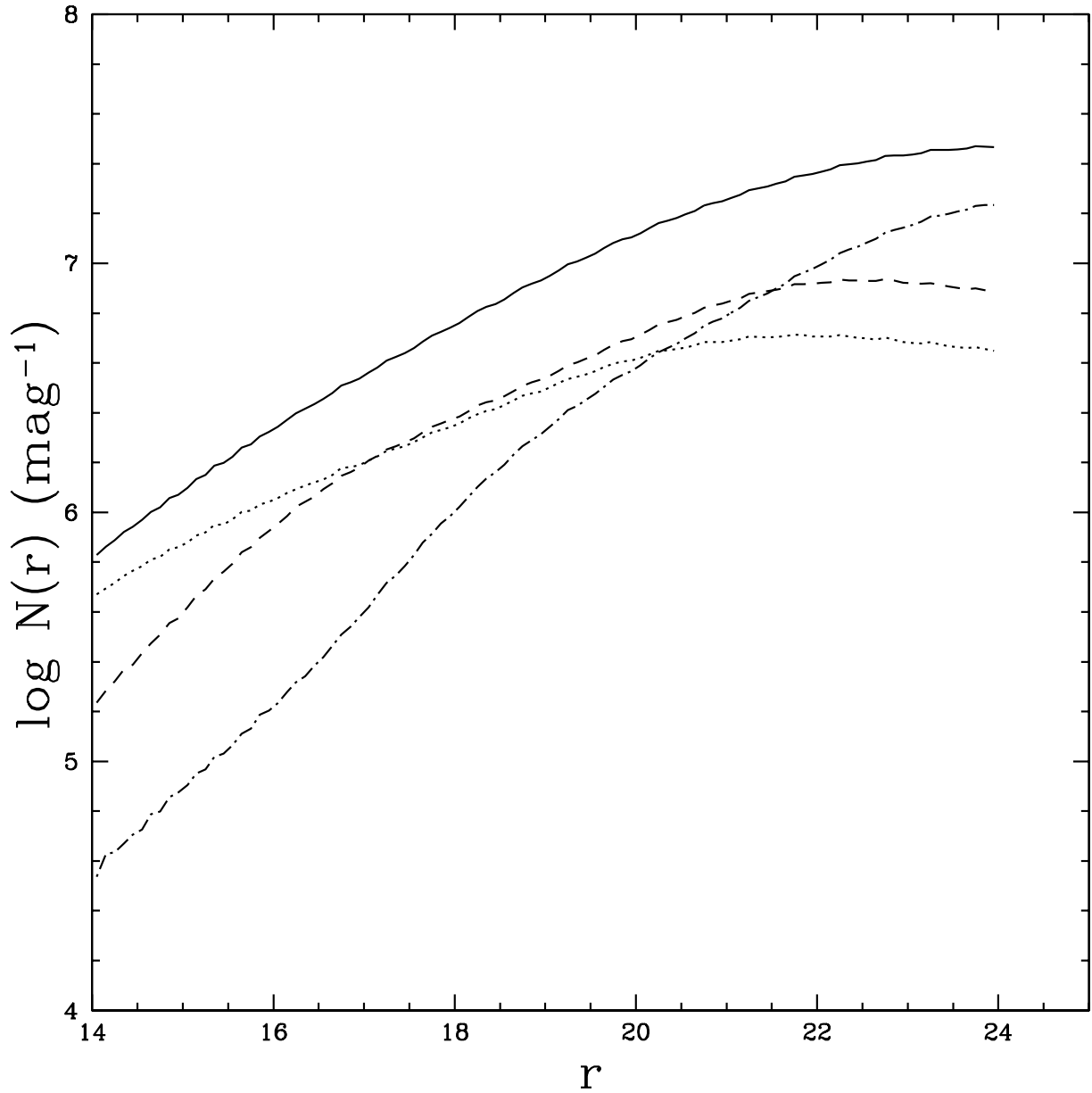}}
\caption{\small Star counts (in $mag^{-1}$) as a function of $r$ magnitude for the entire DES solid angle, according to the model described in \S \ref{models}. The dotted line indicates the contribution from the thin disk, the dashed line shows the thick disk star counts, and the dot-dashed line shows the stellar halo. The solid line gives the total star counts. The adopted bin width is 0.1 mag.}
\label{desmagdist}
\end{figure}

The basic structural parameters used to model the DES data are listed in Table \ref{tabstruct}. The fiducial model parameters for the disks, bulge, and the halo were taken from the aforementioned studies used to validate TRILEGAL itself.

The model also assumes that the Sun is at $R_{\odot}=8500$ pc from the Galactic center and displaced by $z_{\odot}=24 pc$ from the disk mid-plane. Stars are randomly drawn from a piecewise power-law IMF as quoted by \citet{Kroupa2001}. In order to mimic unresolved binarism, a fraction of 50\% of the stars
are assigned a companion star from the same IMF, with a mass ratio also randomly chosen in the range $0.6 \leq m_2 / m_1 \leq 1.0$. Extinction is drawn from the \citet{Schlegel1998} maps. In order to model populations of different ages and metallicities, the model used distinct star formation histories (SFH) and age-metallicity relations for each component \citep[for details, see][]{Girardi2005}.

Figure \ref{spat_dist} shows the on-sky distribution of model stars inside the DES footprint according to spectral type. The projection is shown in equatorial coordinates. Stripe 82 is the narrow region very close to the Equator. The connection region is the square just to the south of it, extending to $\delta = -20^o$. The main gradient in surface density seen in all panels follows Galactic latitude.

Figure \ref{desmagdist} shows the TRILEGAL predictions for the differential number of stars, in logarithmic scale, as a function of $r$ magnitude for the entire DES solid angle. We use a 0.1 mag bin width but express the numbers in unit mag bins. The total counts are shown as a solid line, whereas the contribution from each structural component is shown as indicated in the captions. The total number of model stars for each Galactic component is $3.13\times10^7$ for the thin disk, $4.20\times10^7$ for the thick disk and $4.65\times10^7$ for the halo. The figure shows that DES will probe stars in a magnitude range ($r>21$) well dominated by thick disk and halo counts. The model includes a total of $\simeq 1.2\times10^8$ stars. Given that the stellar halo and even the thick disk have never been probed to such a large volume, the number of stars may increase if an unanticipated population of low-luminosity white dwarfs or M dwarfs are found in these components. Such a population of cool halo dwarfs has in fact been suggested earlier, although it was later contested \citep[][]{Oppenheimer2001, Bergeron2003}.

\begin{figure}[!ht]
\resizebox{\hsize}{!}{\includegraphics{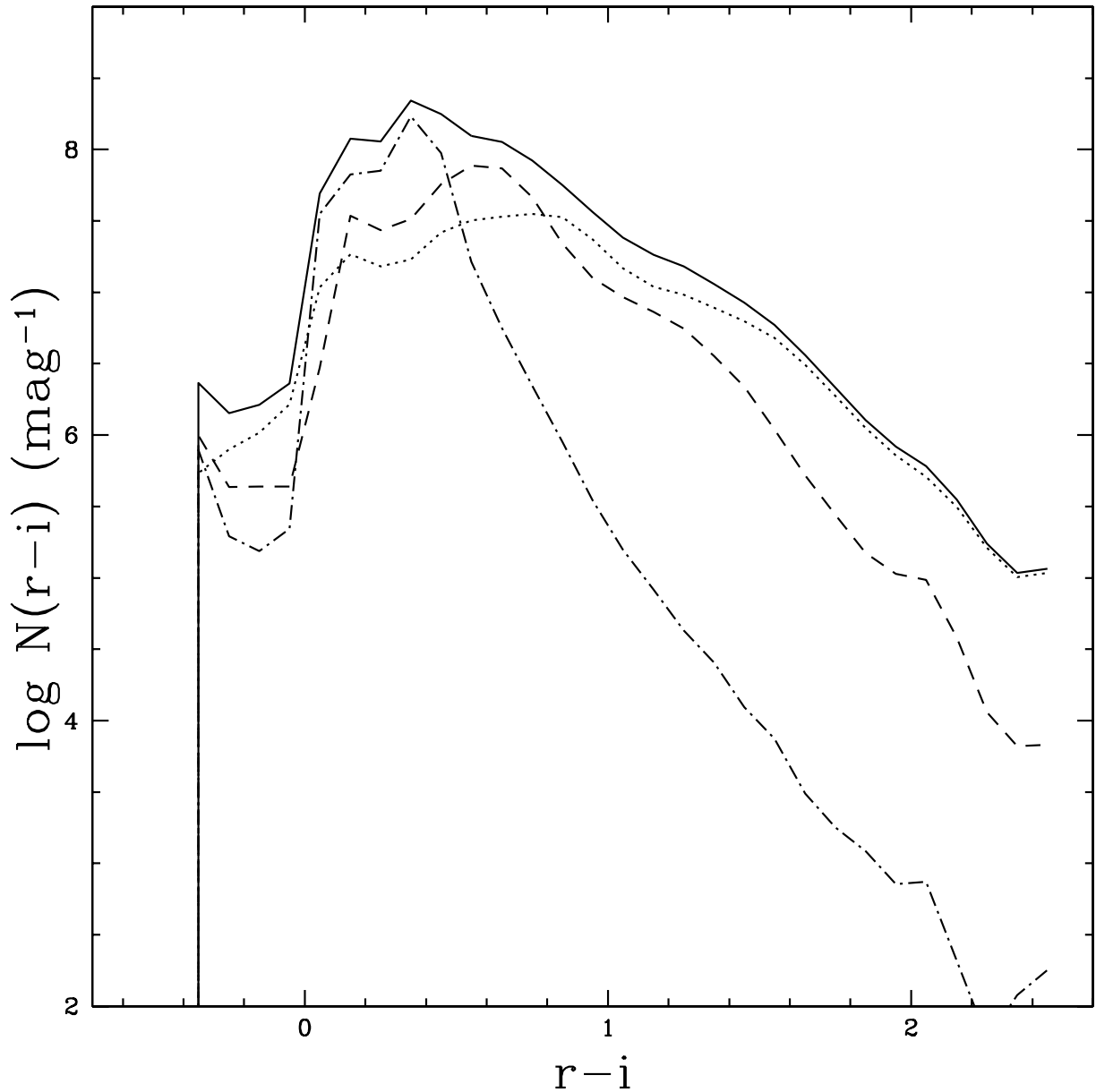}}
\caption{\small Star counts as a function of {\it r-i} color for the entire DES solid angle as predicted by the same model as in Figure \ref{desmagdist}. The binning and line types also follow the same convention as in that figure.}
\label{descoldist}
\end{figure}

The $r-i$ color distribution is shown for the same model in Figure \ref{descoldist}.
The symbols and bin width are the same as in Figure \ref{desmagdist}. The numbers are also normalized to unit magnitude. At the blue end, $r-i \leq 0.6$, the distribution is dominated by halo and thick disk stars close to the main sequence turn-off (MSTO). The thin disk low-mass main sequence contributes with most of the stars in the range $r-i \geq 1$.

\section{Star-galaxy separation}
\label{stargalsep}
The usefulness of DES to studies of the MW structure and resolved stellar populations in general depends strongly on the ability to explore its stellar sample to its full depth. This requires an efficient method to distinguish stars from galaxies and QSOs down to the faint DES detection levels. We here concentrate on star-galaxy separation, since galaxies are by far the main source of contamination to a stellar sample, especially at faint magnitudes ($r \geq 22$).

DES images are simulated as part of DES Data Challenges. These experiments serve as a means to test and improve on the survey's data management pipelines. They also serve the purpose of evaluating the survey requirements (including object classification) and validating science analysis pipelines prior to the availability of real data. Data Challenge images attempt to be as realistic as possible, mimicking
not only the expected geometry of DECam images, but also various instrumental effects, including point spread function variations due to atmospheric seeing and CCD position, combined transmission and response curves from filters, CCDs and telescope, and so on. 

The input stellar sample to the Data Challenges was created with TRILEGAL.
The input galaxy sample was generated by AddGals (M. Busha, 
private communication), which is an algorithm 
for {\it painting} galaxies onto dark matter particles in an N-body simulation 
by matching galaxy luminosities with local dark matter densities, not dark 
matter halos. In addition to an N-body simulation, the 
algorithm takes as inputs a galaxy luminosity function, a luminosity-dependent
 correlation function, and a distribution of galaxy colors given luminosity 
and environment.

\begin{figure}[!ht]
\resizebox{\hsize}{!}{\includegraphics{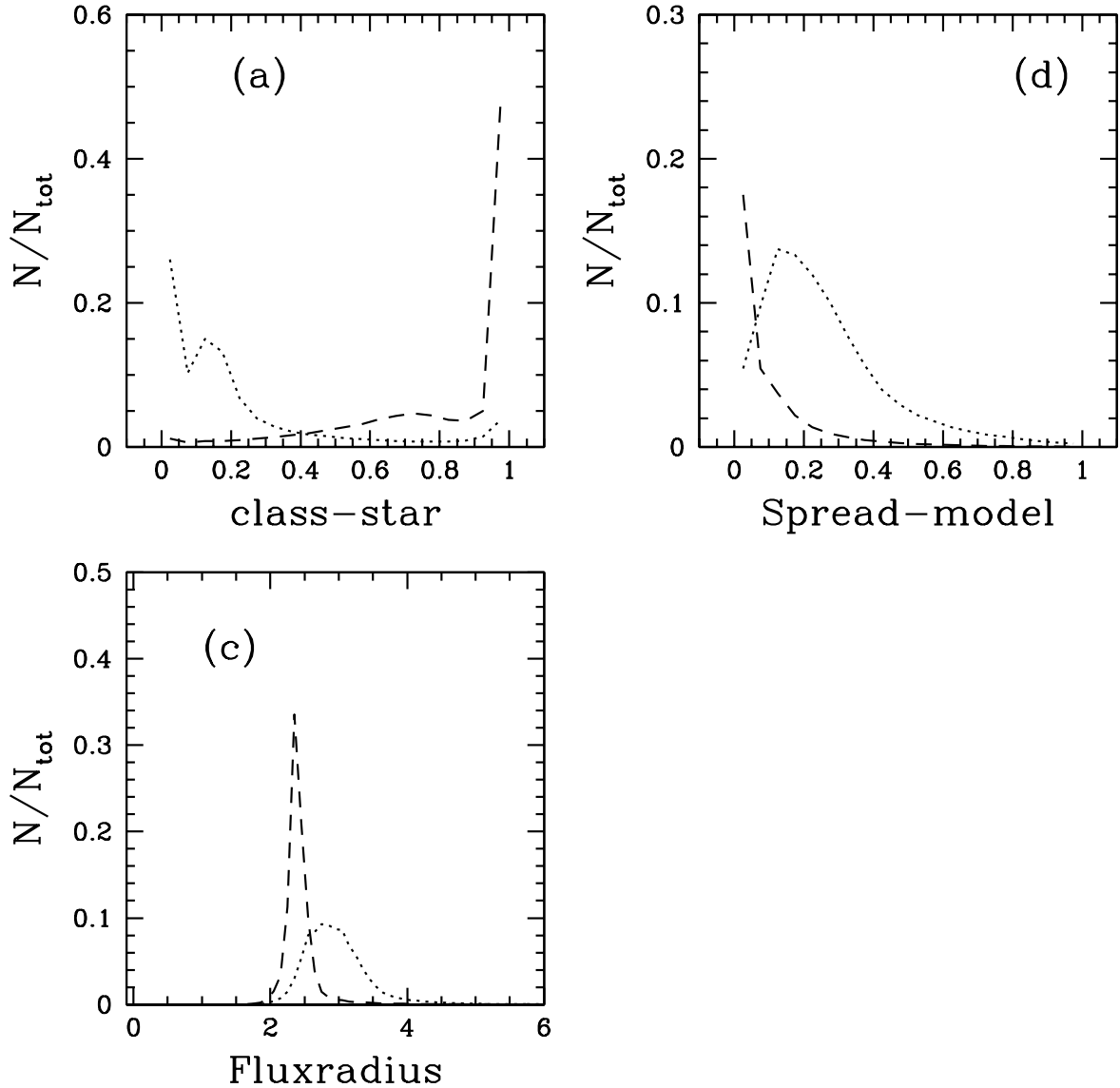}}
\caption{\small Distribution of different image-based classification
  parameters for stars and galaxies resulting from DC5
  simulations. The distributions for stars (galaxies) are shown in
  dashed (dotted) lines.
Panel {\it a)}: distribution of class-star; {\it b)} Spread-model; {\it c)} Fluxradius. }
\label{plparsdist}
\end{figure}

Tests with the latest set of such simulations, from Data Challenge 5 (DC5), have been carried out in order to assess how different star-galaxy classification parameters perform. The distributions of three different such parameters are shown separately for stars and galaxies in Figure \ref {plparsdist}.
DES Data Management pipelines use SExtractor as their basic source detection and characterization tool. SExtractor outputs a stellarity index (SI) or class-star for each object it finds. This parameter is presented in \cite{Bertin96}. Fluxradius is a measure of the half-light radius of each source. Finally, spread-model is a classifier adopted in SDSS \citep{Scranton2002}. A star-galaxy classification scheme depends not only on the adoption of a discriminating parameter but also on some cut-off value for this quantity. In the case of class-star, for instance, galaxies tend to have values close to zero, while the distribution of class-star values for stars concentrates close to unity. A typical cut-off value for selecting stars is class-star$ > 0.9$.

To quantify the star-galaxy separation efficiency of a given parameter and cut-off value, we build
completeness and purity curves. Completeness is defined as the the ratio of the number of true stars classified as such in a given magnitude range to the total number of true stars in that range. Purity is defined as the ratio of the number of true stars classified as such in a given magnitude bin to the total number of objects that were classified as stars in the same bin.

\begin{figure}[!ht]
\resizebox{\hsize}{!}{\includegraphics{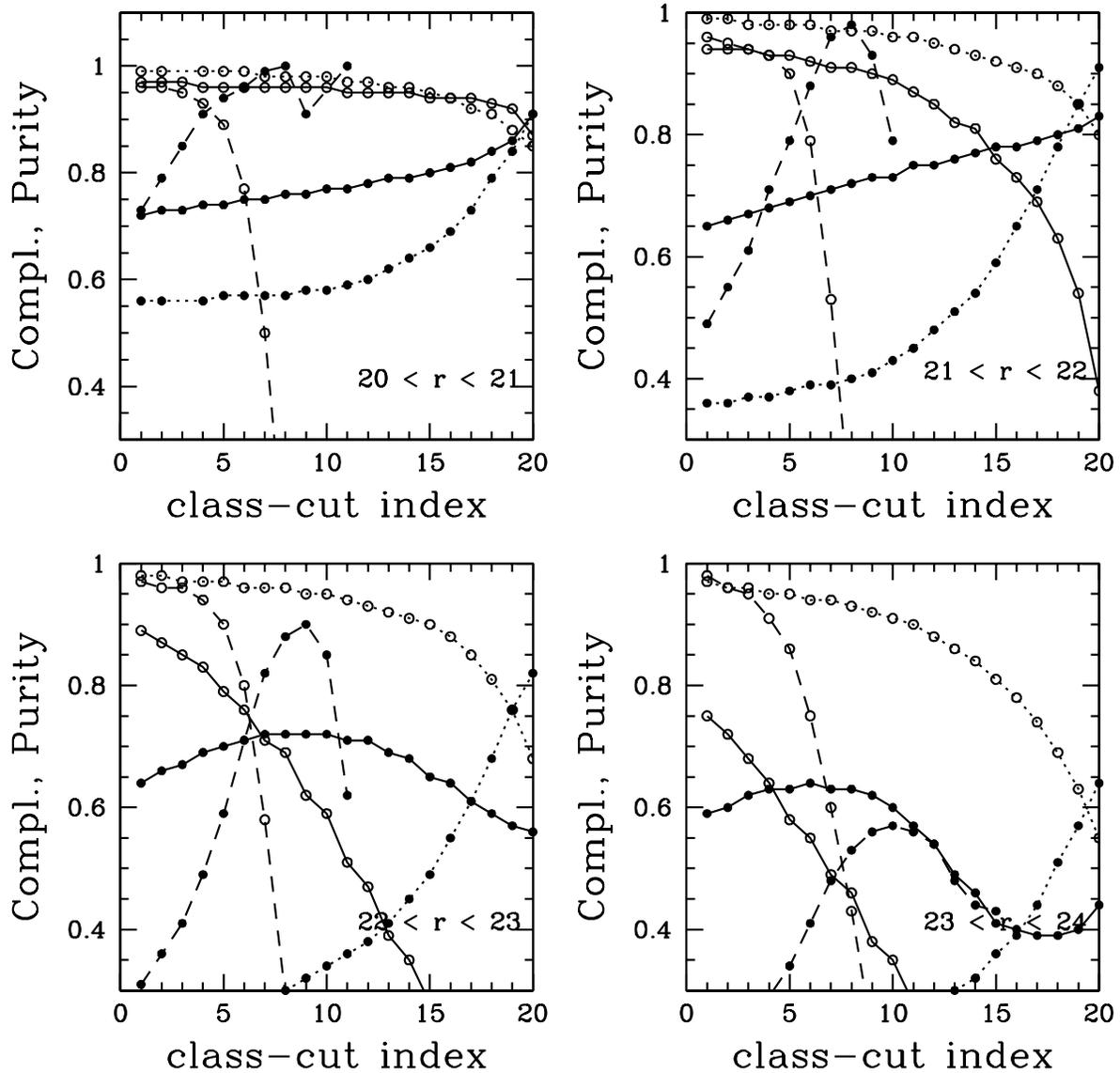}}
\caption{\small Completeness and purity curves are shown in open and
  filled circles, respectively, for different classification parameters and for different magnitude ranges. The magnitude ranges are indicated in each panel. The curve types are the same in all panels: solid: class-star; dashed: fluxradius; dotted: spread-model. In the faintest magnitude bin ($23 < r < 24$; 
lower right panel), notice how class-star (solid lines) leads to a 
simultaneous purity and completeness of 63\%, performing better than spread-model (60\%) and fluxradius (51\%)}
\label{plhist1}
\end{figure}

Completeness and purity curves from DC5 data are shown in Figure
\ref{plhist1} as open and filled circles, respectively. The different panels correspond to different magnitude ranges, as indicated. The solid lines show the results of using class-star as a star-galaxy separator, the dashed lines show the curves based on fluxradius and the dotted ones are based on spread-model. The cut-off index along the x-axis depends on the parameter and its cut-off value used. For class-star, whose domain is in the range $[0,1]$, we define the cut-off index as (class-star$-0.40)/0.03$.
Therefore, an index value of zero means an extreme situation where a selection of class-star$ > 0.4$ was used for the stellar sample. In the case of fluxradius the mapping into cut-off index is given by $(3-$fluxradius$)/0.1$. And for spread-model we use $(0.4 - $spread-model$)/0.02$. These transformations always map the parameters onto the $[0,20]$ range in cut-off index, thus allowing the completeness and purity curves to be plotted together.

We notice from Figure \ref{plhist1} that completeness and purity compete with one another. A loose cut-off index (close to zero) makes the stellar sample complete, but very contaminated. A stringent cut-off, on the other hand, has the opposite effect. The point where these two curves cross indicates a good compromise between the two quantities. The DC5 results shown in the figure indicate that spread-model and fluxradius better discriminate among extended and point sources at bright magnitudes (first 3 panels), whereas class-star performs better at faint magnitude levels. For $r < 22$, a simultaneous value of 80\% in completeness and purity is easily achieved. At the faintest magnitude bin, however, this value drops to 65\%.

An improvement to the object classification scheme stems from using
color information to pre-select a star enriched sample. Color-color
diagrams of stars  and galaxies from the DC5 simulations show that the
stellar locus is better defined in the $(i-z) vs. (r-i)$ panel. We
then pre-select a star enriched subset of the data using this diagram.

\begin{figure}[!ht]
\resizebox{\hsize}{!}{\includegraphics{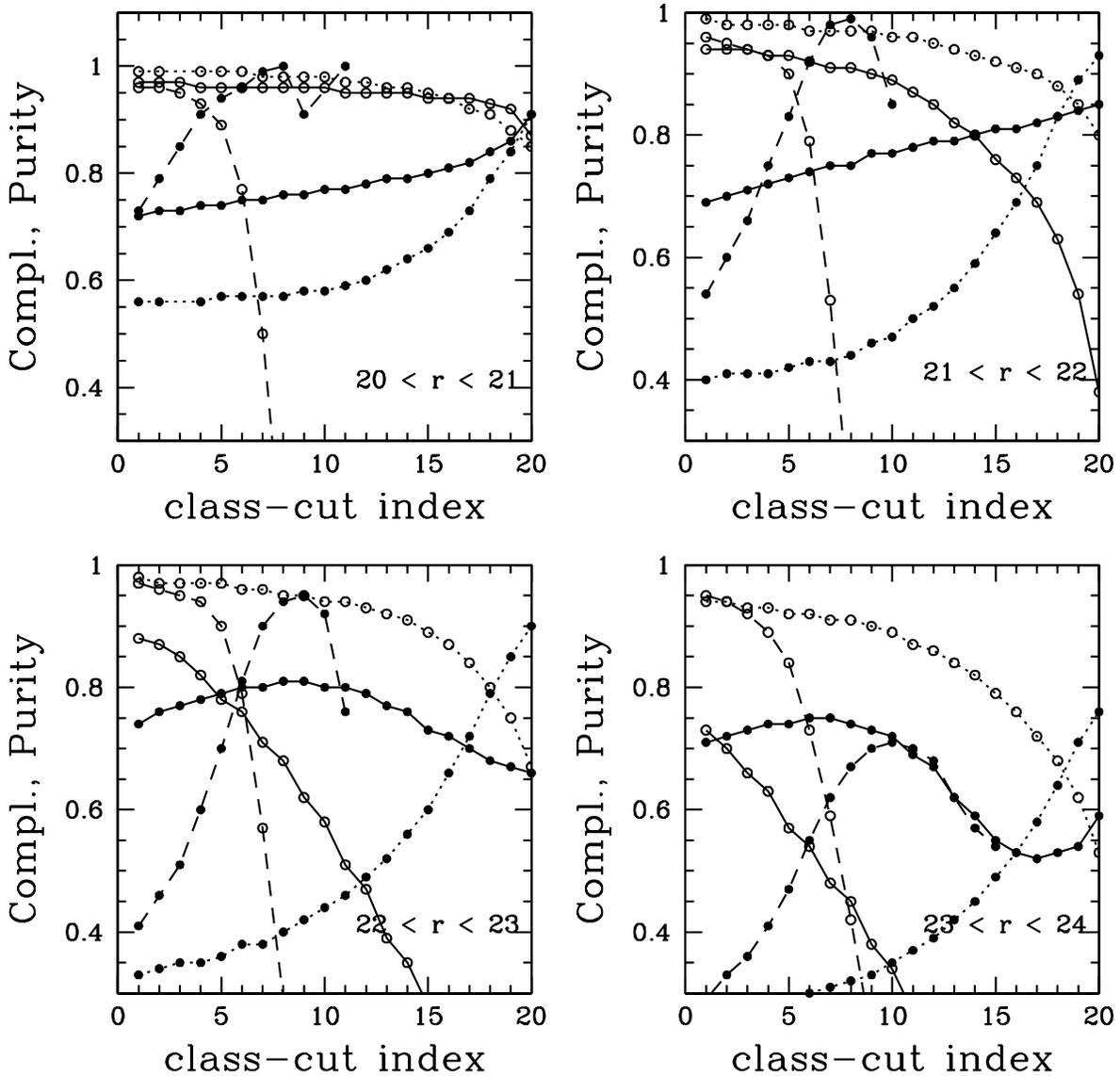}}
\caption{\small Same curves and conventions as in Figure \ref{plhist1}
  but now using a color-based pre-selection meant to boost the
  fraction of stars. Completeness and purity curves are shown as open
  and filled circles, respectively, for different classification parameters and for different magnitude ranges. The magnitude ranges are indicated in each panel. The curve types are the same in all panels: solid:
class-star; dashed: fluxradius; dotted: spread-model.}
\label{plhist2}
\end{figure}

Figure \ref{plhist2} shows completeness and purity curves for the same 3 object classifiers as in Figure \ref{plhist1} but now applied to the color pre-selected sample. Since a large chunk of the galaxy sample was eliminated from the sample based on colors, the purity values are boosted relative to the previous figure. The overall behavior of the curves remains but now an 80\% pure and complete stellar sample is obtained down to $r=23$. Class-star still does a better job in separating stars from galaxies at low S/N levels, leading to a complete and pure stellar sample at the 72\% level.

The star-galaxy separation experiments we carried out so far do not yet take into account all the color information that may be used to classify the sources. Another alternative, which we are currently exploring, is to abandon the classification paradigm and use these image parameters, along with colors, magnitude and positions to assign probabilities of objects being stars, galaxies and QSOs. In what follows, however, we will conservatively assume that the stars in our Galactic model will suffer from a 30\% incompleteness and an equal amount of contamination by extra-galactic sources.

\section{The Stellar and Sub-stellar Luminosity Functions}
\label{stellar}
Given that stars can be successfully separated from galaxies and QSOs down to the nominal survey
photometric depth, DES will provide stellar samples several times deeper than SDSS. The deep DES images will sample a large number of main sequence M stars, whose mass is slightly above or at the hydrogen burning limit, imposing stronger constraints than previous data on the very faint end of the thick disk and halo stellar luminosity functions. DES will also significantly increase the census of sub-stellar objects such as L and T dwarfs, as well as of white dwarfs close to or at the end of the cooling
sequence. Table \ref{tabdepths} below lists these interesting stellar types whose luminosity function is still relatively unconstrained at the faint end ($M_V > 13$), specially in the thick disk and halo components. The DES depth limits were calculated by taking into account the survey limits in each filter and the minimum requirement that at least one color index is needed to select candidates. The values listed do not take extinction into account.
\medskip
\placetable{tabdepths}
\begin{table}
\begin{center}
\caption{\small DES sampling depths for M, L, T, and white dwarfs.\label{tabdepths}}
\resizebox{0.47\textwidth}{!} {
\begin{tabular}{ll@{}@{}lr}
\tableline\tableline
Type & \multicolumn{2}{c}{Abs. Mag.} & maximum distance \\
\tableline
M0 dwarf & $M_i = 8.4$; & $M_z = 8.0 $ & 13.3 kpc \\
M9 dwarf & $M_i = 15.5$; & $M_z = 14.0$ & 500 pc \\
L5 BD & $M_i = 18.0$; & $M_z = 15.8$ & 160 pc \\
T0 BD & $M_i = 20.0$; & $M_z = 17.5$ & 65 pc \\
T5 BD & $M_i = 23.0$; & $M_z = 19.0$ & 16 pc \\
pre LF peak WD & $M_g = 13.0$; & $M_i = 13.0$ & 900 pc \\
post LF peak WD & $M_g = 16.5$; & $M_i = 15.1$ & 180 pc \\
\tableline
\end{tabular}
}
\end{center}
\end{table}
\subsection {Cool low-mass stars}
\label{coolstellar}
DES depth limits for M, L or T dwarfs are based on the min(i,z) detection limits. The absolute magnitudes were taken from \citet{Hawley2002}. The depth values computed in this way may actually be an underestimate, since an alternative approach to pre-select such extremely cool objects is to find objects detected in $z$ (or $i$ and $z$) but which drop out at bluer wavelengths. Overall, the
estimated DES depths are from 3 to 8 times larger than those from SDSS as estimated by \citet{Hawley2002}.

\begin{figure}[p]
  \begin{minipage}{0.450\textwidth}
    \resizebox{\hsize}{!}{\includegraphics{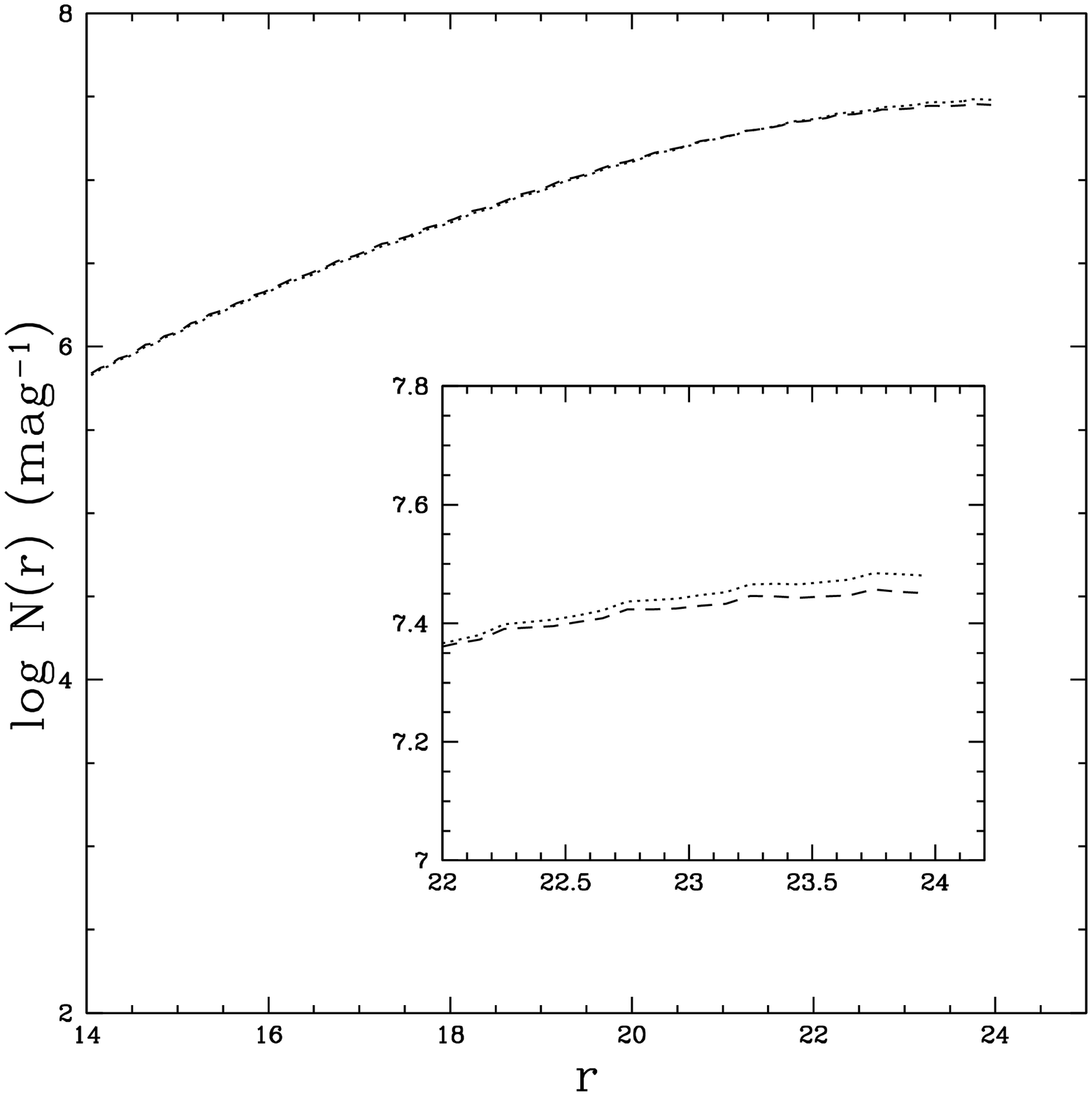}}
  \end{minipage}
  \\
  \begin{minipage}{0.450\textwidth}
    \resizebox{\textwidth}{!}{\includegraphics{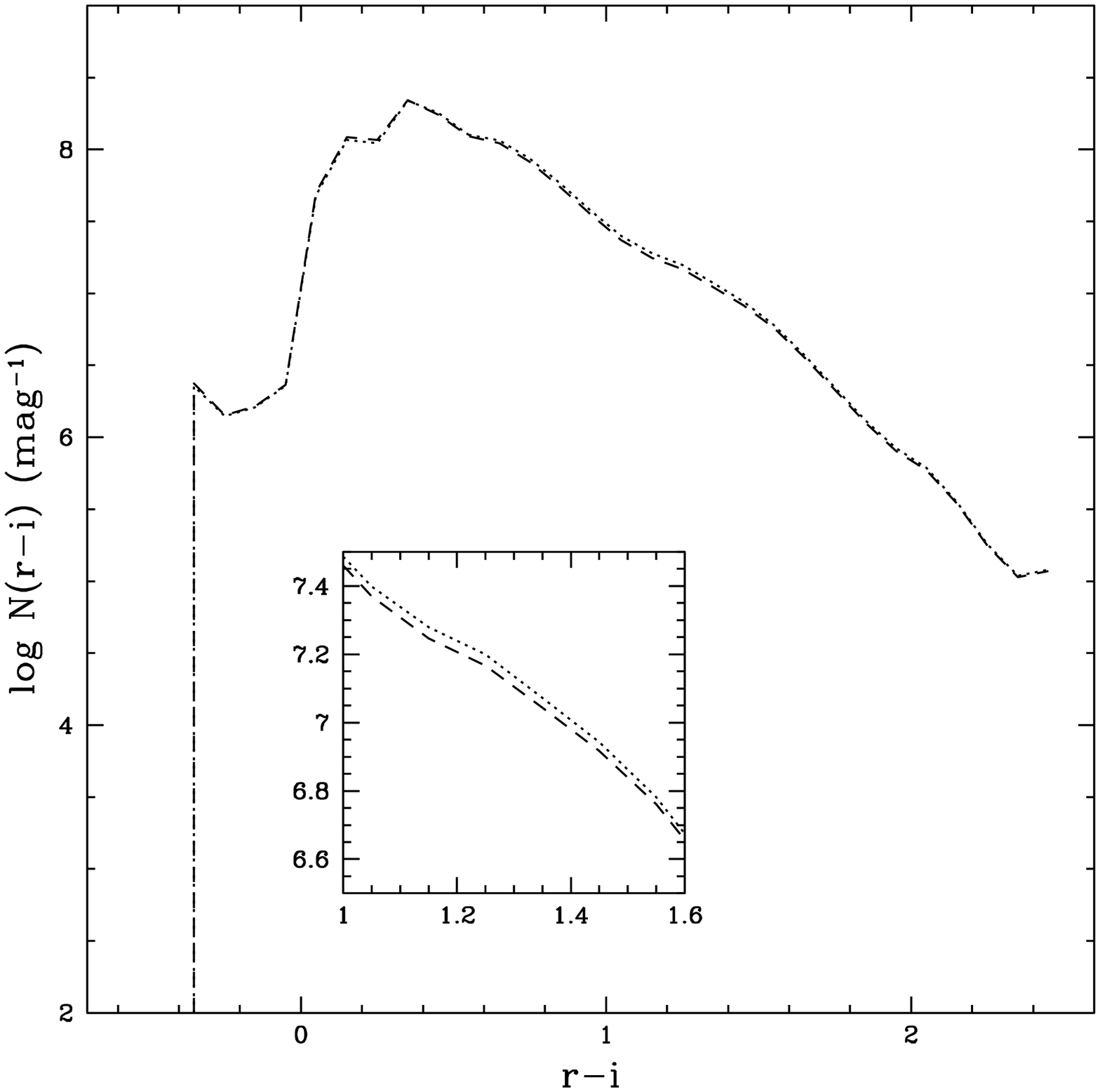}}
  \end{minipage}
  \caption{\small Upper panel: Model star counts as a function of $r$ magnitude for the entire DES solid angle under two assumptions concerning the slope of the IMF in the $0.08 - 0.50~M_{\odot}$ mass range for the halo and thick disk. The bin size is 0.1 mag. The dotted (dashed) line corresponds to an IMF slope, $d \log \phi (m) / d \log m = 1.4$ ($d \log \phi (m) / d \log m =1.2$) in the quoted range. The inset at the bottom shows a blow up of the plot fainter than $r=22$, where the low-mass star counts contribution is larger. Lower panel: Model star counts as a function of $r-i$ color for the entire DES solid angle for the same models as in the previous panel. The inset at the bottom again shows a blow up of the color range where the star counts differ most.}
  \label{msslopemstars}
\end{figure}

The larger depths available to DES will allow strong constraints on the slope of the thick disk and stellar halo IMF close to the hydrogen burning mass limit. In Figure \ref{msslopemstars} we show the expected number counts in the entire DES footprint based on two slightly different assumptions for the IMF slope of the thick disk and halo populations in the mass interval $0.08 < m/m_{\odot} < 0.50$. We use the \citet{Kroupa2001} IMF as a starting point, whose power-law slope in this mass range is $d \log\phi(m) / d \log m = 1.3$. The dotted line shows the results based on the fiducial model (\S \ref{models}) but with the slope $d \log\phi(m) / d \log m = 1.4$. The dashed line shows the results for $d \log\phi(m) / d \log m = 1.2$. The difference in star counts caused by these subtle variations in the fiducial IMF slope is larger at the faint end of the magnitude distribution and at the $1.0 \leq r-i \leq 1.6$ color range. This color range is typical of low-metallicity M dwarfs.
In absolute terms, the number counts in the quoted mass range above is of $\simeq 7.5\times10^7$ stars in the entire DES footprint, using the reference IMF slope, and including thin disk stars. This is more than 50\% of all DES stars. The difference in star counts shown in Figure \ref{msslopemstars} is of $\simeq 3\times10^6$, which is only about 4\% of the total counts of M and K dwarfs in the fiducial model. Still this difference is three orders of magnitude larger than the associated Poisson fluctuations in the total number count.

This very large statistical significance resulting from subtle variations in the model IMF slope is not strongly affected if we incorporate the star/galaxy separation efficiency discussed in \S \ref{stargalsep}. If we conservatively assume a 50\% photometric incompleteness for K and M dwarfs in the stellar halo on top of a 30\% incompleteness caused by star/galaxy separation, the difference in star counts will drop by a factor of three, to $\simeq 10^6$ stars. In addition, if we now contaminate the K and M stellar sample by 30\% of galaxies, the total number of $0.08 < m/m_{\odot} < 0.50$ stars in the fiducial model increases to $\simeq 10^8$. This means that the expected variation in star counts caused by the IMF change is still $\simeq 10^6 / 10^4 = 10^2$ times larger than the Poisson fluctuations the total model counts.
We thus conclude that the observed DES star counts, especially in the color range quoted above, can be safely used to constrain the halo and thick disk IMF slope at low-masses, with high statistical significance, to within a few hundredths dex.

We carried out a similar exercise, but now varying the IMF slope in the sub-stellar regime, $0.01 < m/m_{\odot} < 0.08$. Our fiducial IMF predicts a power-law slope of 0.3 dex in this range for all Galactic components. This leads to $\simeq 3\times10^4$ brown dwarfs in our fiducial model, almost all of them belonging to the thin disk.  An IMF  variation of 0.2 dex leads to a 10\% change in the number counts. Incorporating a factor of 3 decrease due to incompleteness leads to a difference of $\simeq 10^3$ in the number counts of these very red objects when the IMF is varied. This is three orders of magnitude smaller a difference in star counts than in the previous experiment. The expected Poisson fluctuations in the fiducial model will be of the order of $\sqrt{1.3\times3\times10^4}=100$, where the $1.3$ factor again accounts for galaxy contamination. Notice that this contamination will probably be smaller at the very red colors typical of L and T dwarfs (see below). Still these overestimated fluctuations are one order of magnitude smaller than the expected variation in the numbers of these objects. We may thus conclude that the census of very red faint objects in DES may be used to constrain the IMF slope of sub-stellar objects to the precision of about 0.1 dex.

It is useful to compare the color range of M, L, and T dwarfs to the expected colors of galaxies, since these latter will dominate the DES source counts at faint magnitudes. \citet{Newberg2002} analyzed
about 5 million stars from SDSS close to the celestial Equator. The observed color-magnitude diagrams (CMDs) show that galaxies strongly dominate the source counts at relatively blue color ranges $0.4 < g-r < 1.0$, being much less numerous in the $g-r > 1.1$ range occupied by low-mass main sequence stars or by sub-stellar objects. We are currently in the process of quantifying the purity of the expected DES stellar sample when it is selected by a combination of stellarity parameters from the images and observed colors. Our goal is to build a probability distribution function of a given DES source to be a star as a function of position, magnitude and color.
\subsection {White dwarfs}
\label{wdstellar}
\begin{figure}[!ht]
\resizebox{\hsize}{!}{\includegraphics{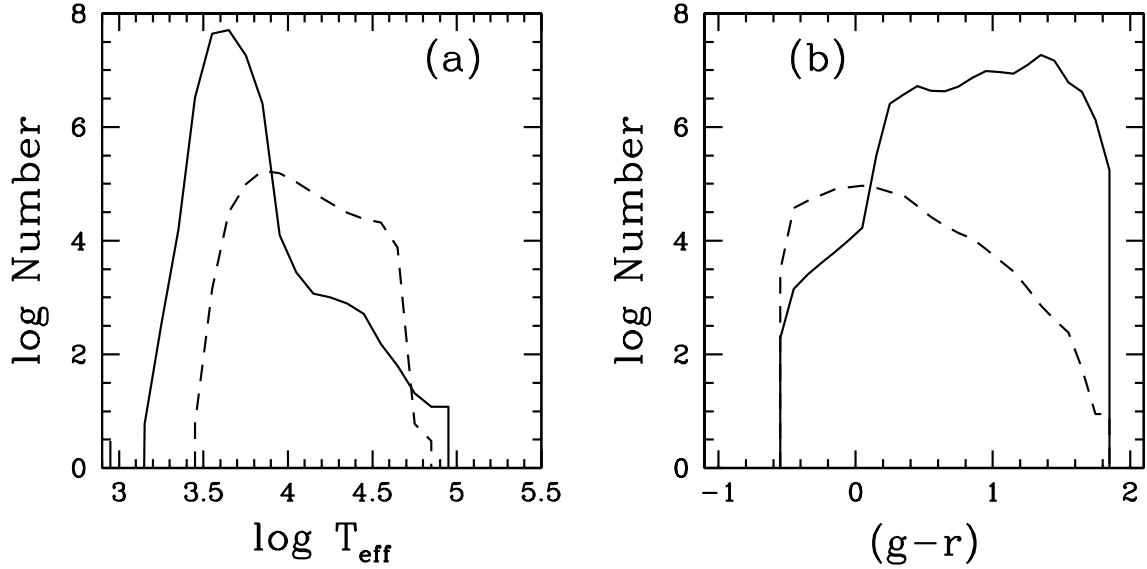}}
\caption{\small Panel {\it a)} distribution of $T_{eff}$ values of main sequence stars ($\log g < 6$; solid line) and white dwarfs ($\log g > 6.5$ and $\log L/L_{\odot} < 0$; dashed line). Panel {\it b)} Distribution of $(g-r)$ colors
using the same line conventions as in the previous panel.}
\label{plwds}
\end{figure}

For white dwarfs (WD), the DES depths shown in Table \ref{tabdepths} were computed using the min(g,i)
detection distance. Absolute magnitudes are from \citet{Bergeron1995} models for a purely hydrogen WD with $\log g = 8$. We here assess the usefulness of DES filters to separate a sample of WDs from the other stellar types. White dwarfs have much larger surface gravities and tend to be relatively
blue compared to the bulk of the main sequence stars. In Figure \ref{plwds} we compare the properties of WDs with those of normal stars in our fiducial Galactic model, described in \S \ref{models}.
We do that both in the theoretical and in the observational space. Panel {\it a} shows the distribution of effective temperatures ($T_{eff}$) of main sequence
stars and white dwarfs, separated according to $\log g$.
This quantity clearly separates main sequence stars
($\log g < 6$) from the WD cooling sequence ($\log g > 6.5$ and $log L/L_{\odot} < 0$). The total number of WDs in our model is $7.6\times10^5$.
The distributions as a function of $(g-r)$ colours are shown in panel {\it b}.
Notice that white dwarfs outnumber main sequence stars by over an order of magnitude for $T_{eff} > 10,000$K. However, the distributions are not so clearly segregated in color. Still, a cut in $(g-r) < 0$ will tend to select a sample of white dwarfs with 20\% contamination or less.
The results from our simulations are in qualitative agreement with previous work based on SDSS colors by \citet{Harris2003}, where it is shown that WDs hotter than $\simeq 12,000$K segregate in color
space from normal main sequence stars, BHB and QSOs. However, availability of the $u$-band is particularly useful to derive a cleaner sample of white dwarfs and to separate them from BHBs and QSOs. Follow up imaging with the DECam or availability of SkyMapper data will provide $u$-band magnitudes.

\section {Proper motions}
\label{proper}
Cool WDs are mixed with and largely outnumbered by halo stars close to the MSTO. As these WDs probe the faint end of the WDLF, they are extremely important as a tool to date the thin and thick disks at different scale heights. The best means to disentangle such a cool WD population is to use proper motions (PMs), since disk WDs, being at much shorter distances, will display larger PMs than the distant F and G halo stars \citep{Kilic2006}. Assuming a typical transverse velocity of $V_T = 20$ km s$^{-1}$relative to the local standard of rest and a distance of $d=100$ pc, leads to $\mu \simeq 40$ mas yr$^{-1}$. At larger distances perpendicular to the disk plane, similar values may be expected since the velocity dispersion also tends to increase. A cool halo WD at $d=500$ pc with $V_T = 200$ km s$^{-1}$ will have $\mu \simeq 80$ mas yr$^{-1}$. Such proper motions are likely to be measurable over the 5 year baseline of DES.

Proper motions will also be useful to separate BDs from high redshift QSOs and very red galaxies. The expected PM values of BDs, which will be detected well within the thin disk (see Table \ref{tabdepths}) are similar to those of cool disk WDs, since they share similar kinematics and distances.

Given their importance to the selection of interesting stellar types, we have implemented and validated a kinematical model in the TRILEGAL code. Space velocities are randomly taken from the Schwarzschild velocity distribution appropriate for the age, position, and galactic component of each simulated star. The velocity ellipsoids are taken from the
literature, and assumed to be cyllindrically-simmetric with respect to the Galactic disk rotation axis\footnote{This is probably a good approximation for the disk populations that will be observed by DES, but can certainly be improved for the stellar halo.}. For the thin disk, the UVW velocities are taken from \citet{Holmberg2009}, with an extrapolation for the oldest ages and truncated for the youngest, and suitably converted into the cyllindrical $V_r, V_{\phi}, V_z$. The local
$V_{\phi}$ is then inreased/reduced as a function of the galactocentric radius, according to the galactic rotation curve from \citet{Dias2005}. For the thick disk and halo, the velocity ellipsoids are taken from \citet{Chiba2000}.

Space velocities generated this way are then corrected by the motion with respect to the local standard of rest (from \citet{Dehnen1998}), and decomposed into radial and transverse components (see \citet{Johnson1987}). The latter is finally converted into proper motions using the simulated stellar distance.

\begin{figure*}
  \begin{minipage}{0.450\textwidth}
    \resizebox{\hsize}{!}{\includegraphics[angle=-90]{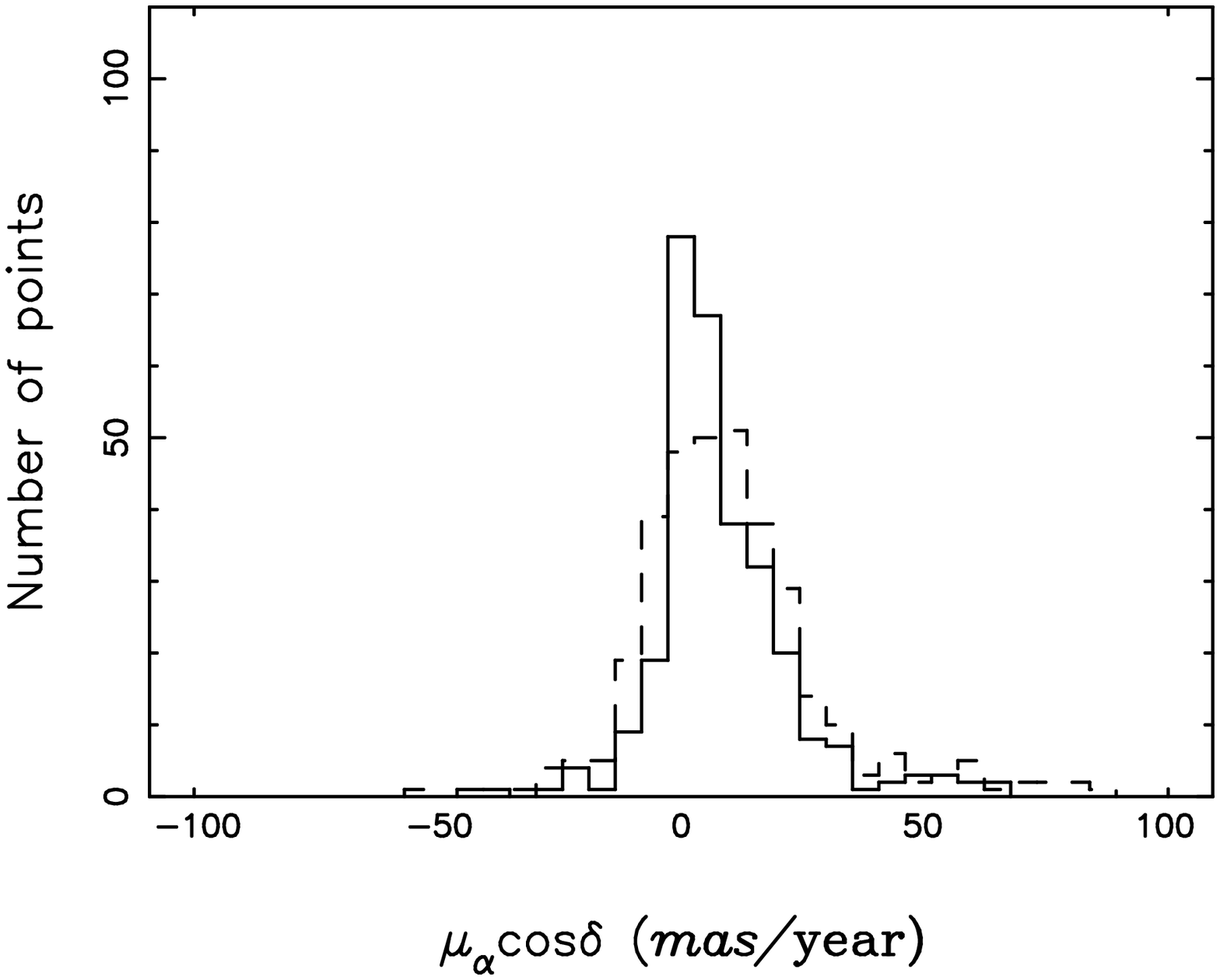}}
  \end{minipage}
  \begin{minipage}{0.450\textwidth}
    \resizebox{\hsize}{!}{\includegraphics[angle=-90]{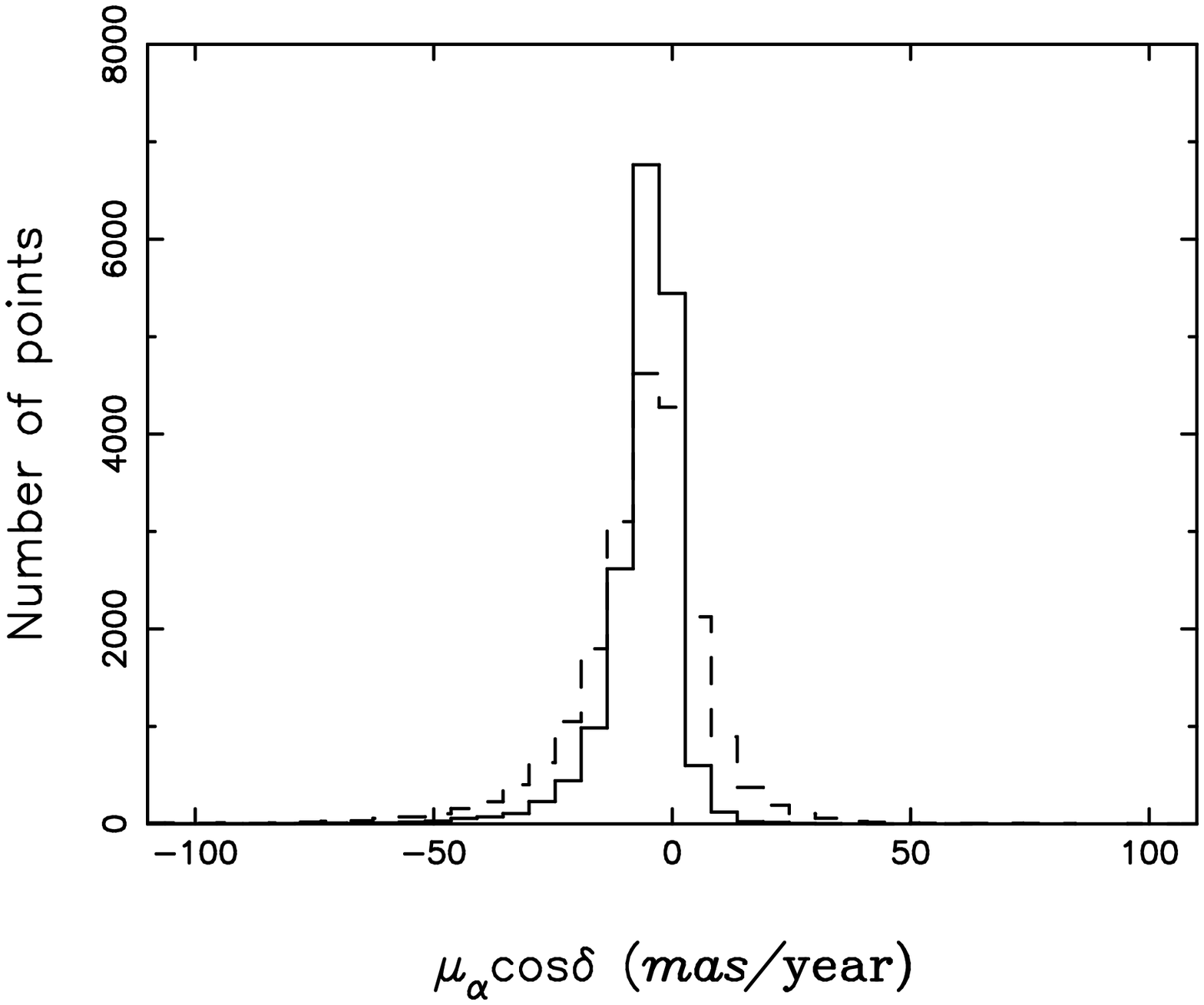}}
  \end{minipage}
  \\
  \begin{minipage}{0.450\textwidth}
    \resizebox{\hsize}{!}{\includegraphics[angle=-90]{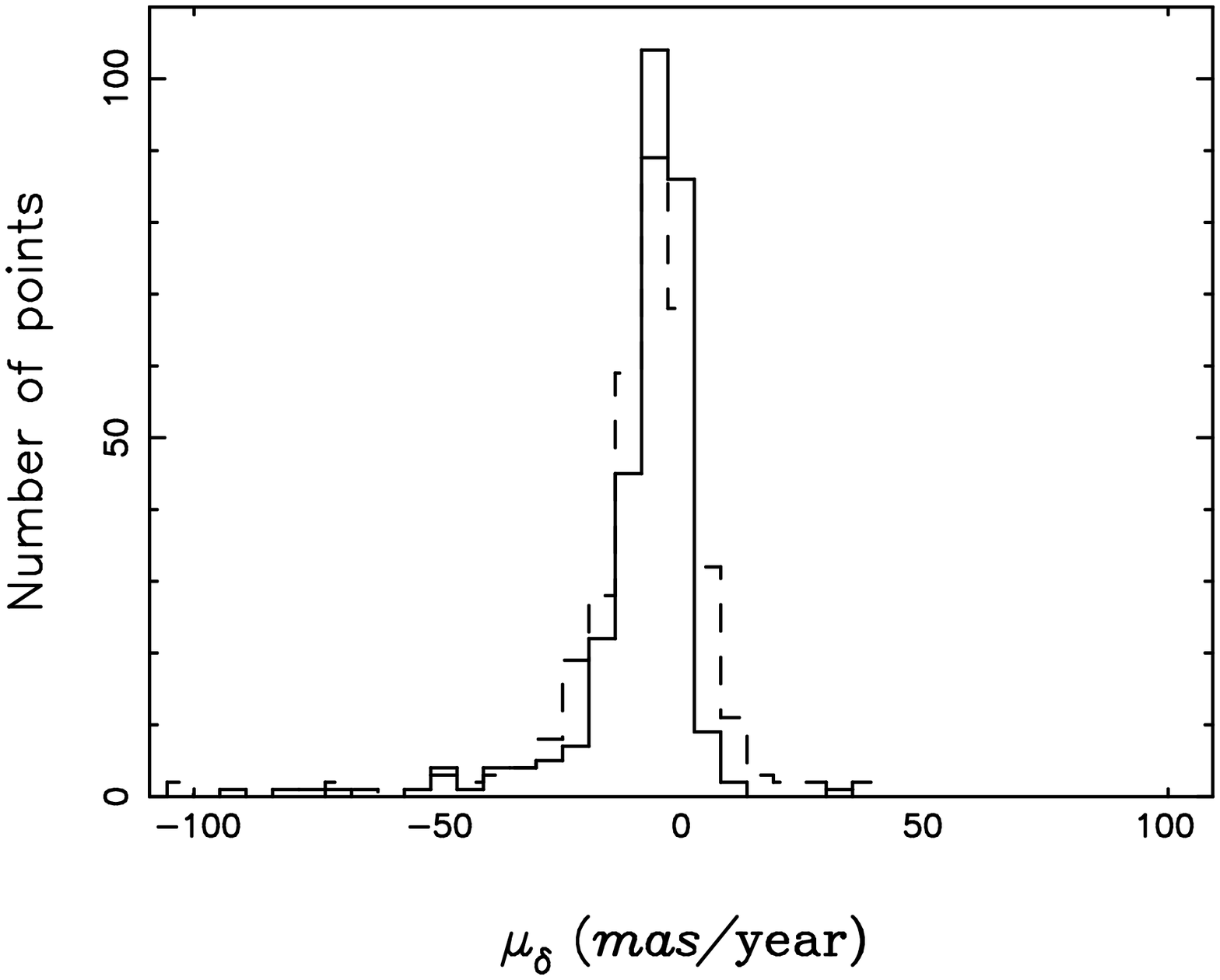}}
  \end{minipage}
  \begin{minipage}{0.450\textwidth}
    \resizebox{\hsize}{!}{\includegraphics[angle=-90]{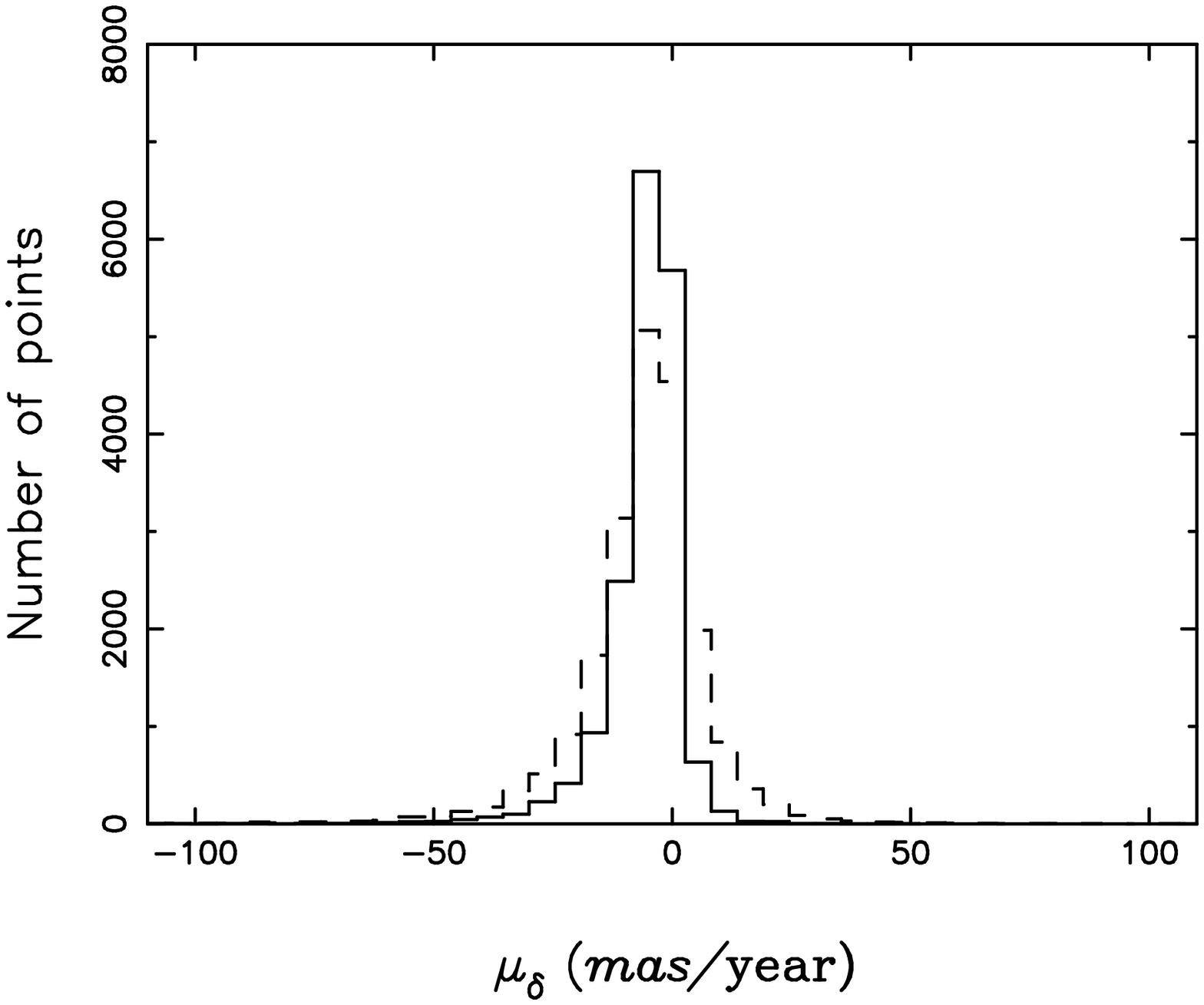}}
  \end{minipage}
  \caption{\small Comparison of the TRILEGAL proper motion distributions (solid line) with the observed ones (dashed line). The upper panels show $\mu_{\alpha} cos \delta$ distributions, whereas the lower panels show those in $\mu_{\delta}$. In the left (right) panels the model is compared to UCAC3 (SDSS) data. The model was cut at $V=17$ for the UCAC3 comparison and at $g=22$ for SDSS in an attempt to simulate the typical depth of these reference data.}
  \label{triucac}
\end{figure*}

In Figure \ref{triucac}, we compare the model PMs with data.  The panels on the left show the results
for PMs from the 3rd version of the U.S. Naval Observatory CCD Astrometric Catalog \citep[UCAC3,][]{Zacharias2010}. The field is a 1 deg$^2$ region centered on $(l,b)=(352.5,-62.8)$, which is inside the region of the DC5 simulations, described in \S \ref{stargalsep}. The TRILEGAL data were cut at bright ($V < 17$) magnitudes in order to match UCAC3 depth.
The model and observed PM distributions are very similar. The narrower peak in the model can be accounted for by the measurement errors which are not yet simulated. Notice that the spread in proper motions is larger parallel to the Equator than perpendicular to it. The panels on the right of Figure \ref{triucac} show the model comparison to SDSS PMs in a mirror field to the first one relative to the disk plane, $(l,b)=(352.5,+62.8)$. In this case the model stars were cut at $g < 22$, again as an attempt to approximately mimic the SDSS sample. The model and data PM distributions, both parallel and perpendicular to the Equator, are again nearly identical apart from the lack of model errors. The TRILEGAL kinematical mode was also validated in other directions, including close to the Galactic and
Equatorial Poles, the Solar Apex, and directions close to the Galactic center and anti-center. In all these directions the model PMs reproduced UCAC3 distributions very well.
We also investigated the dependence of the PM values with magnitude for a field in the DES DC5 region and found that the PM distributions change very little as a function of magnitude. 

We are thus in a position to simulate the expected distributions of PMs within the entire DES footprint and for different stellar types, including WDs and BDs. These predictions will then be compared with the observed ones resulting from DES astrometry. It is likely that the best PM estimated for faint stars such as WDs and BDs will result from DES itself. The reason is the lack of earlier data sets with comparable depth in the Southern Hemisphere. The current astrometric precision aimed by DES data management is of $\simeq 50$ mas for the 5-year stacks. The survey strategy is still not entirely
defined, so it is currently not possible to couple the astrometric precision with a time baseline and in order to assess the expected PM errors.
\section{Disk and Halo Structure}
\label{halostruct}
\begin{figure}
  \begin{minipage}{0.450\textwidth}
    \resizebox{\hsize}{!}{\includegraphics{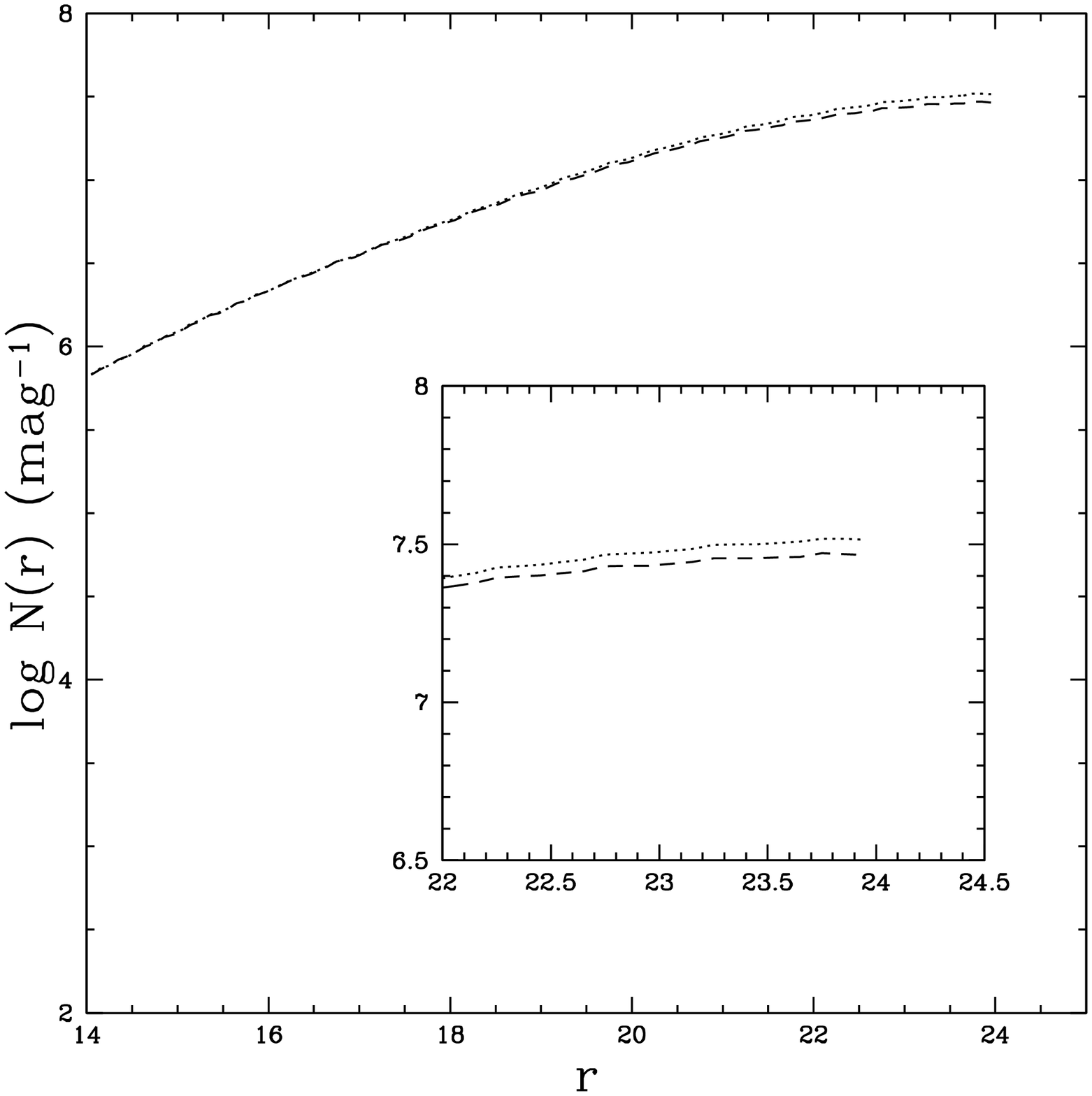}}
  \end{minipage}
  \\
  \begin{minipage}{0.450\textwidth}
    \resizebox{\hsize}{!}{\includegraphics{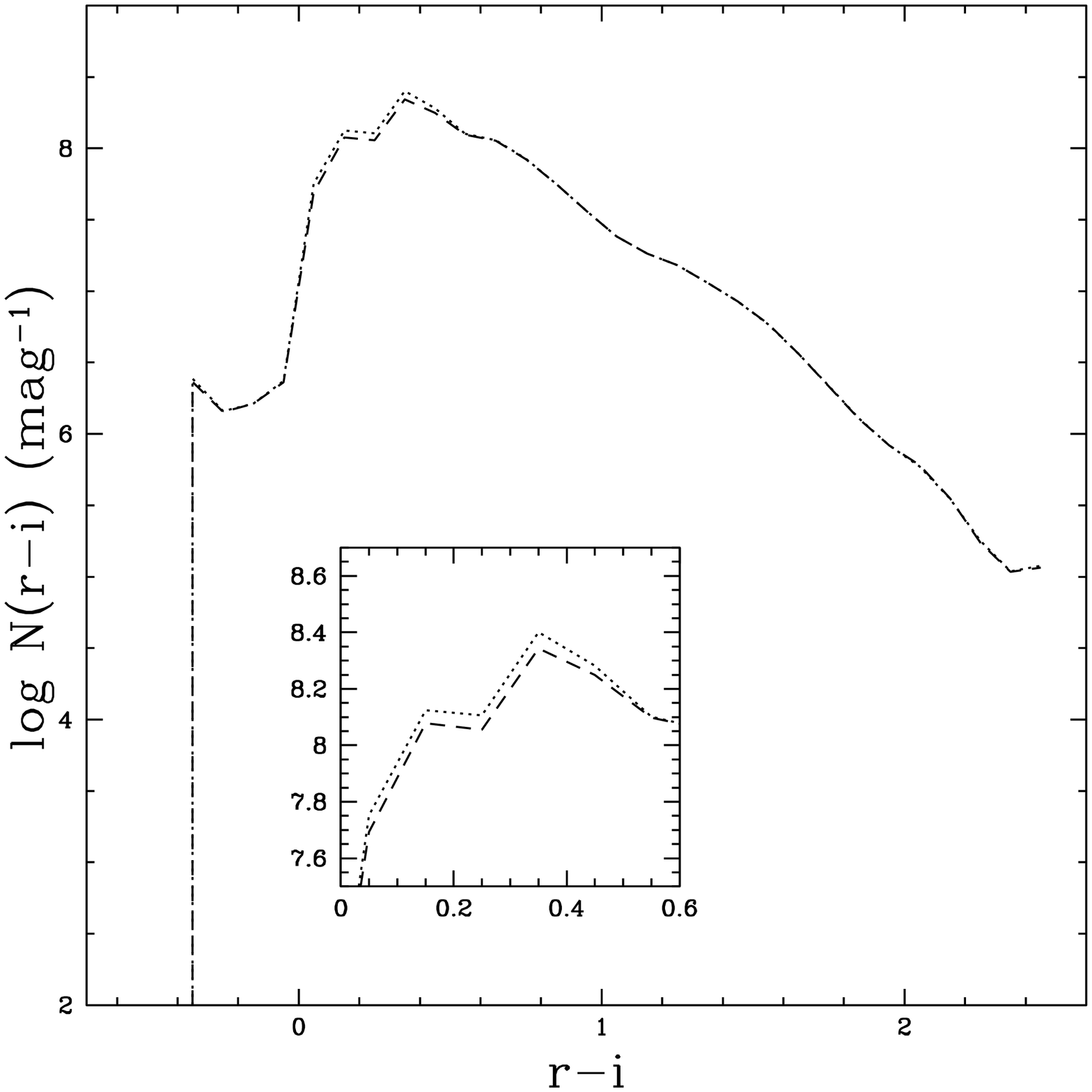}}
  \end{minipage}
  \caption{\small Upper panel: Model star counts as a function of $r$ magnitude in the entire DES solid angle for two different choices of an oblate halo axis ratio, $c/a$. dotted line: $c/a = 0.7$; dashed line: $c/a = 0.65$. The bin size is 0.1 mag. The inset at the bottom right is a blow up in the range $r > 22$, where the differences are more pronounced. Lower panel: Star counts as a function of $r-i$ color for the same models as in the previous panel. The inset at the bottom right is a blow up in the color range where the differences are more pronounced.}
  \label{haloshape}
\end{figure}
Models of the structure of the Galaxy require a large area coverage in order to quantify the various parameters governing the density profile and geometry of the different structural components. The basic input for the tomography of the Milky Way are the star counts as a function of magnitude, colors and direction on the sky. Strong constraints on the model parameters can be established using photometric parallaxes of stellar tracers of each Galactic component. A recent and extensive work on this line is that of \citet{Juric2008}. The authors derived scale lengths, scale heights and relative
normalization for the thin and thick disk components using the distribution of low mass main sequence stars (M dwarfs) sampled by SDSS out to $\simeq 2$ kpc. The structure of the stellar halo, on the other hand, was constrained using F and G stars close to the its MSTO.

In an alternative approach, \citet{deJong2010} applied a CMD fitting technique to the SDSS Extension for Galactic Understanding and Exploration \citep[SEGUE,][]{Yanny2009} sample in order to constrain the structure of the thick disk and stellar halo. This technique models the Galactic components at once by building artificial CMDs at different directions and statistically comparing them to the observed ones.

One point of discrepancy among different authors is the shape of the stellar halo. \citet{deJong2010} favor an oblate halo with an axis ratio of $c/a=0.88$ within the inner 25 kpc. \citet{Juric2008} best fitting models are also oblate but flatter: $c/a = 0.64$. Finally, \citet{Du2008} find a varying stellar halo shape with evidence of a lower $c/a \simeq 0.4$ towards the center of the Galaxy and $c/a \simeq 0.6$ in the opposite direction.

Figure \ref{haloshape} shows predicted star counts for DES with 2 different assumptions for an oblate halo axial ratio: $c/a=0.65$ (which is our reference value) and $c/a=0.70$. The difference between the two models becomes clearer for $r > 21.0$, where halo counts start to become significative compared to the other components. But, again, the best bet to constrain the halo structure is to use color counts, in this case in the range $0 < r-i < 0.6$, which is dominated by F and G stars close to the halo MSTO. About $4\times10^7$ stars are found in this color range, while the difference in number counts between the two models is at the 10\% level. As in the case of the M dwarfs, this variation is extremely significant from a statistical point of view. Galaxy contamination will be significant in the
color domain of these halo MSTO stars, rendering the constraints to the halo shape more sensitive to our capability to separate DES stars from galaxies down to the survey's photometric limits. Incorporating the effects of our conservative estimate of 70\% classification purity and completeness, plus 50\% photometric incompleteness, we expect the two slightly different halo shapes to yield
a difference of $\simeq 10^6$ in observed star counts, which is to be compared with the $\simeq \sqrt{1.3\times4\times10^7} \simeq 7\times10^3$ Poisson fluctuation in these number counts.

DES will allow comparison of star counts between the Southern and Northern hemispheres to a precision and volume not previously obtainable. \citet{Xu2007} noted evidence for a significant asymmetry in halo stars counts between North (using SDSS) and South. Unfortunately, the work relied on use of photographic plate data in the South, and it was difficult to compare to the deeper, more precise CCD data in the North. DES will offer a vastly improved CCD data set in the $griz$ filters in the South,
which we will explore in detail, and compare against models. Similarly, some recent studies also discuss a possible triaxiality in the thick disk \citep[][]{Parker2003,Parker2004,Larsen2010}, which again will be better investigated with a combined sample covering both hemispheres.

Any global asymmetry found in the structure of the stellar halo can be used as input into Galactic formation models, and opens the door to quantitatively explore triaxial dark matter halos, as well as a possible misalignment between a dark matter halo and the Milky Way stellar disk.

\subsection{Halo Sub-structure}
\label{halosubstruct}
Several substructures are expected to exist and have been found in the more extended Galactic components, especially in the halo. In particular, the currently most successful scenario for structure formation in the universe, the $\Lambda$CDM model, predicts that large galaxies form by the agglutination of sub-galactic fragments, specially during their early evolutionary history.
Consistently with this picture, numerous new and unexpected streams of stars have been detected in the
Northern Galactic Cap using the SDSS data.

The Sagittarius tidal stream is the most prominent structure, running from (l,b) = (200$^o$,30$^o$) to (l, b) = (340$^o$,60$^o$). The Sagittarius tidal stream is also prominent in the South, and crosses through the DES survey area. Also visible are the Orphan Stream, running perpendicularly from (l,b) = (255$^o$,48$^o$) to (l,b) = (195$^o$,48$^o$), and the Virgo over density at (l, b) = (300$^o$,60$^o$), among others.

DES will be able to carry out a similar survey and detect streams in the South to even fainter depth. The fundamental physics obtainable from modeling the orbits of these streams are constraints on the shape of the dark matter halo in which the tracer stream stars move. Work on streams in the north
indicates that several streams at different inclinations are needed to probe the potential accurately.

New satellites of the Galaxy will also be discovered with the stellar DES sample. SDSS has yielded a total of 14 new dwarfs spheroidals or transition objects between low-mass galaxies and globular clusters \citep[see][and references therein]{Walsh2009}. An analysis by \citet{Tollerud2008} results in an estimate between $19$ and $37$ new satellites to be discovered in DES assuming that the characteristics of detectability are the same as in SDSS.
On the other hand, the deeper DES sampling should increase the contrast of these objects relative to
halo field stars, since DES will probe further down the stellar luminosity function. In particular, DES will be able to probe the stellar main sequence of an old population at distances out to $\simeq 120$ kpc. This will lead to a much faster increase in the number of satellite stars as compared to the contaminating background. Assuming a power-law IMF, $dN/dm \propto m^{-\alpha}$ and a simple power-law MS mass-luminosity relation $L \propto m^{\beta}$, we can integrate the IMF in order to estimate the number of sampled stars in a given population. The lower limit of the integral is
the lowest MS mass which leads to detection (given the satellite distance and the sample magnitude limit). The upper limit will be the highest MS mass which has not yet turned into a degenerate (which decreases with age) The relative increase in the number counts of stars for two surveys with magnitude limits $m_1$ and $m_2 > m_1$ is then

$${ {\Delta N_*} \over {N_*} } = { {10^{-0.4{{1-\alpha}\over{\beta}}m_2} -10^{-0.4{{1-\alpha}\over{\beta}}m_1} }
\over {10^{-0.4{{1-\alpha}\over{\beta}}m_1} - 10^{-0.4{{1-\alpha}\over{\beta}}m_{br}} } },$$

where $m_{br}$ in the denominator is the MS magnitude of the brightest star in the population. If we assume that this second term on the denominator is zero ($m_1 >> m_{br}$, which is the same as saying that the number of degenerates is insignificant) and use $\alpha = 2.35$ (Salpeter IMF), $\beta=3.5$, $m_1 = 22.5$ (SDSS) and $m_2 = 24$ (DES), one obtains $\Delta N_* / N_* = 0.7$. In other words, considering that the background counts of stars vary much more slowly, DES will increase the density contrast of MW satellites by a factor of 1.7 relative to SDSS out to a distance modulus of $(m-M)_0 \simeq 20$. This factor will be larger if we relax the assumption that the satellite has no degenerates.
We thus conclude that DES will be particularly more efficient than SDSS in detecting very low-mass satellites at distances about twice as large as those to the Magellanic Clouds.

Detection of satellites situated further away will not be as significantly improved unless they have young or intermediate age stellar populations, for which the previous argument would apply to a more luminous MS locus. One interesting possibility that also needs to be further investigated is any dependence of typical satellite size with distance from the MW. A trend of smaller clusters having more concentrated spatial distributions has been observed in extragalactic cluster systems and may as well
apply to MW satellite galaxies and transition objects \citep[][]{Larsen03, Jordan05}. This would make the more distant MW satellites of lower surface brightness at a fixed mass than their more nearby counterparts, something that would jeopardize their detection.
We are currently in the process of investigating the detection of MW satellites with DES in more detail, by taking selection effects with surface brightness and star formation history into account.
\section{Summary and Conclusions}
\label{summary}
In this paper we have utilized a theoretical model to forecast the main contributions that DES will give to studies of Galactic structure and resolved stellar populations. We adopted a fiducial Galactic
model based on the TRILEGAL code, as described in \S \ref{models}. It predicts $\simeq 1.2\times10^8$ stars in total within the DES footprint. Thin disk star counts will be dominated by low mass main sequence stars with $m \leq 0.5 m_{\odot}$. But such low mass stars will also be sampled in large numbers in the thick disk and halo, and beyond about 10 kpc. Consequently, we have shown that the IMF slope at low masses will be constrained to within a few hundredths dex in the thick disk and halo.

About $3\times10^4$ L and T dwarfs will also be sampled by DES, the vast majority of them belonging to the thin disk. This will again allow a strong constraint to the IMF in the sub-stellar regime.
We estimate that the sampling depth of low mass and low-luminosity species, such as M, L, and T dwarfs or white dwarfs, will be from 3 to 8 times larger than those quoted for SDSS in the literature.
The Galactic model we adopted was also used to discuss the capacity of detection of white dwarfs using $grizY$ colors. We conclude that a sample of hot white dwarfs, with a contamination of about 20\% or less, can be selected without the $u$-band.
We also present the initial results of our model proper motions and compare them to observed ones from
UCAC3 and SDSS towards several directions in the Galaxy. In all cases the model proper motions agree with the data. We argue in favor of using the proper motions determined from DES as a means to detect cool white dwarfs and to separate brown dwarfs from high redshift QSOs.

We have also addressed one of the main issues for stellar studies in a deep photometric survey such as DES: star-galaxy separation. We used the latest set of DES image and catalogue simulations
to quantify completeness and purity of a stellar sample as selected with different image classification parameters. We conclude that a global stellar sample which is simultaneously pure and complete at the 70\% will easily be achieved. We also discussed the effect of the remaining contamination on the science results. More specific stellar samples may be further refined on basis of the multi-dimensional color space available from DES. One example is the use of color information in association with image based stellarity parameters, which will likely allow red stellar and sub-stellar objects to be more efficiently separated from most galaxies.

Thick disk and halo stars will be dominated by F and G stars close to the MSTO point. One of the main contributions from these objects will be to probe the halo structure and sub-structure to very large
distances. Our models show that DES star counts will constrain the halo shape to very high accuracy. For instance, an oblate halo will have its axial ratio determined to within $1-2\%$, assuming
an efficient star/galaxy separation down $r \simeq 23.5$. Asymmetries in star counts as a function of direction on the sky and magnitude, including those based on a joint DES/SDSS sample, will constrain halo triaxiality as well as thick disk azimuthal variations.

Finally, we discuss the impact that DES will have on the detection of new halo substructure, including new star clusters and low-mass galaxies. We conclude that DES will increase the density contrast of these objects by about 60\% or 70\% when compared to SDSS for distances lower than $\simeq 120$ kpc. This is the result of DES reaching the stellar main-sequence of an old stellar population within this distance range. This enhanced efficiency distance limit may increase for satellites that harbor young or intermediate age stellar populations.
%

\acknowledgments{ {\bf Acknowledgments.}

Funding for the DES Projects has been provided by the U.S. Department of Energy, the U.S. National Science Foundation, the Ministry of Science and Education of Spain, the Science and Technology Facilities Council of the United Kingdom, the Higher Education Funding Council for England, the National Center for Supercomputing Applications at the University of Illinois at Urbana-Champaign, the Kavli Institute of Cosmological Physics at the University of Chicago, Financiadora de Estudos e Projetos, Funda\c c\~ao Carlos Chagas Filho de Amparo \`a Pesquisa do Estado do Rio de Janeiro,
Conselho Nacional de Desenvolvimento Cient\'\i fico e Tecnol\'ogico and the Minist\'erio da Ci\^encia e Tecnologia, the Deutsche Forschungsgemeinschaft and the Collaborating Institutions in the Dark Energy
Survey.

The Collaborating Institutions are Argonne National Laboratories, the University of California at Santa Cruz, the University of Cambridge, Centro de Investigaciones Energeticas, Medioambientales y Tecnologicas-Madrid, the University of Chicago, University College London, DES-Brazil, Fermilab,
the University of Edinburgh, the University of Illinois at Urbana-Champaign, the Institut de Ciencies de l'Espai (IEEC/CSIC), the Institut de Fisica d'Altes Energies, the Lawrence Berkeley National Laboratory, the Ludwig-Maximilians Universitat and the associated Excellence Cluster Universe,
the University of Michigan, the National Optical Astronomy Observatory, the University of Nottingham, the Ohio State University, the University of Pennsylvania, the University of Portsmouth, SLAC, Stanford University, the University of Sussex, and Texas A\&M University.

The authors thank the entire DES team for useful discussions and active support. We are grateful to Helio Rocha-Pinto for his contribution to the TRILEGAL kinematical model. We also thank Eduardo Am\^ores for his early contribution and useful discussions. This work was only possible with the infra-structure provided by the DES-Brazil consortium, to which
most authors belong. This research was made possible by the Laborat\'orio Interinstitucional de e-Astronomia (LIneA) operated jointly by the Centro Brasileiro de Pesquisas F\'\i sicas (CBPF), the Laborat\'orio Nacional de Computa\c c\~ao Cient\'\i fica (LNCC) and the Observat\'orio Nacional (ON) and funded by the Ministry of Science and Technology (MCT)
}


%

%

%
\end{document}